\documentclass[fleqn,usenatbib]{mnras}

\usepackage{newtxtext,newtxmath}

\usepackage[T1]{fontenc}
\usepackage{ae,aecompl}

\usepackage{graphicx}	
\usepackage{supertabular}
\usepackage[caption=false]{subfig}
\usepackage{multirow}
\usepackage{amsmath}	
\usepackage{mathtools}

\usepackage{amssymb}	
\renewcommand{\farcs}{\mbox{\ensuremath{.\!\!^{\prime\prime}}}}
\newcommand{\sdss}{SDSS~J0924+0219}

\usepackage[dvipsnames]{xcolor}
\newcommand{\rev}[1]{{\color{Black}{#1}}}

\newcommand{\kkmspc}{K$\,$km$\,$s$^{-1}$pc$^2$}

\title[SDSS J0924+0219: VLA and ALMA observations]{VLA and ALMA observations of the lensed radio-quiet quasar SDSS J0924+0219: a molecular structure in a \rev{3}$\upmu$Jy radio source}

\author[Badole et al.]{
Shruti Badole$^{1}$\thanks{E-mail: shruti.badole@postgrad.manchester.ac.uk},
Neal Jackson$^{1}$,
Philippa Hartley$^{1,2}$,
Dominique Sluse$^{3}$,
\newauthor
Hannah Stacey$^{4,5,6}$,
H\'ector Vives-Arias$^{7,8}$\\
$^{1}$Jodrell Bank Centre for Astrophysics, Department of Physics \& Astronomy, University of Manchester, Alan Turing Building, \\ Oxford Road, Manchester M13 9PL, UK\\
$^{2}$Square Kilometre Array Organisation, Macclesfield, Cheshire SK11 9DL, UK\\
$^{3}$STAR Institute, Quartier Agora - All\'ee du six Ao\^ut, 19c B-4000 Li\`ege, Belgium\\
$^{4}$ASTRON, Netherlands Institute for Radio Astronomy, Oude Hoogeveensedijk 4, 7991 PD, Dwingeloo, The Netherlands\\
$^{5}$Kapteyn Astronomical Institute, PO Box 800, 9700 AV Groningen, The Netherlands\\
$^{6}$Max Planck Institute for Astrophysics, Karl-Schwarzschild Str. 1, D-85748 Garching bei M\"unchen, Germany\\
$^{7}$Instituto de Astrof\'isica de Canarias, V\'ia L\'actea, s/n, E-38205 La Laguna, Tenerife, Spain\\
$^{8}$Departmento de Astrof\'isica, Universidad de La Laguna, E-38205, La Laguna, Tenerife, Spain\\
}

\date{Accepted 2020 May 21. Received 2020 May 21; in original form 2019 October 23}

\pubyear{2020}
\begin{document}
\label{firstpage}
\pagerange{\pageref{firstpage}--\pageref{lastpage}}
\maketitle
\begin{abstract}
We present Karl G. Jansky Very Large Array (VLA) and Atacama Large Millimetre Array (ALMA) observations of \sdss, a $z$=1.524 radio-quiet lensed quasar with an intrinsic radio flux density of about \rev{3}~$\upmu$Jy. The four lensed images are clearly detected in the radio continuum and the CO(5-4) line, whose centroid is at $z=1.5254\pm0.0001$, with a marginal detection in the submillimetre continuum. The molecular gas displays ordered motion, in a structure approximately \rev{1--2.5~kpc} in physical extent, with typical velocities of 50-100~km$\,$s$^{-1}$. Our results are consistent with the radio emission being emitted from the same region, but not with a point source of radio emission. \sdss\ shows an extreme anomaly in the flux ratios of the two merging images in the optical continuum and broad emission lines, suggesting the influence of microlensing by stars in the lensing galaxy. We find the flux ratio in the \rev{radio, submillimetre continuum and CO lines to be slightly greater than 1 but much less than that in the optical}, which can be reproduced with a smooth galaxy \rev{mass} model and an extended source. Our results, \rev{supported by a microlensing simulation}, suggest that the most likely explanation for the optical flux anomaly is indeed microlensing.
\end{abstract}




\section{Introduction}

\bigskip

\small

In a strong gravitational lensing system, the light from a background source is deflected due to the gravitational potential of a foreground galaxy or galaxy cluster, resulting in multiple \rev{images} of the background source. \rev{The position and brightness of the lensed images} contain information about the structure of the source, together with the mass distribution of the lensing galaxy \citep{Treu2010}. Lensing galaxies or galaxy clusters also magnify the sources by factors of typically 10--20, allowing us to study lensed sources which would otherwise be unobservably faint \citep{Zheng2012ABang, Yuan2012TheRedshift,Atek2015NewA2744}.

%

The use of effective increases in resolution and/or sensitivity is particularly interesting in sources which contain a quasar. About 10\% of quasars are radio loud, with the radio synchrotron emission being generated by jets ejected from close to the central black hole. The remainder generally have weak radio emission, which may be a result of jet activity similar to that in radio loud quasars but on a smaller scale \citep{zak16,her16,whi17,har19,Jarvis2019PrevalenceQuasars}, or due to processes associated with star-formation \citep{con13,bon15,pad17,Stacey2019LoTSS/HETDEX:Quasars}. One way to investigate this question is to study the molecular gas component associated with star-formation processes; this can be detected using molecular lines in the sub-millimetre part of the spectrum, and its morphology and extent compared with the radio continuum. The question of the interaction between AGN and star-forming components is important because it has been suggested for some time (e.g. \citealt{cro06}) that AGN are responsible for regulating star-formation in galaxies, periodically clearing out gas in order to stop star formation. More complex regulatory interactions \rev{have also been proposed} \citep{mai17}. 

The study of molecular gas in lensed submillimetre galaxies has a long history (for recent examples see e.g. \citealt{Hezaveh2013AlmaGalaxies,ryb15,ara16,yan17,gea18}). Molecular gas \rev{reservoirs} have also been found in quasars \citep{Riechers2009Imaging08279+5255,Tuan-Anh2017OnJ0911.4+0551,Paraficz2018ALMAGalaxy}. We can use lens models to correct for the lensing magnification and derive the intrinsic CO luminosity in these objects, and \rev{also} estimate the molecular gas mass if we assume the quantity $\alpha_{\rm CO}$, which relates CO to overall molecular gas mass, is similar to more directly accessible objects ( \citealt{ven03,Riechers2009Imaging08279+5255,rie11,dea13,Tuan-Anh2017OnJ0911.4+0551}).  More recently, detailed imaging together with lens inversion has been used to calculate values for $\alpha_{\rm CO}$ and reconstruct lensed sources in more detail \citep{Paraficz2018ALMAGalaxy,spi19}.

Imaging of radio structures in lensed radio-quiet quasars has also been recently developed, with a number of detections of spatially-resolved radio emission \citep{wuc08,jac11,jac15,har19}. In cases such as RXJ~1131$-$1231, high resolution radio observations resolve out the emission, suggesting emission from an extended disk \citep{wuc08}; in others, variability or directly imaged core-jet structures with VLBI suggest AGN origin as in HS~0810+2554 \citep{har19}. The latter is a surprising result because the object lies within the scatter of the far-infrared (FIR) -- radio correlation 
resulting from the comparison of lensed radio and {\em Herschel} far-infrared luminosities in a large sample of lensed quasars \citep{sta18}, where both radio and FIR emission are expected to be produced by the same, star-formation, mechanism. There is thus a strong motivation to investigate in detail both radio and sub-millimetre emission in the same objects, to derive clues on the relation between radio-emitting components and the molecular gas.

For smooth lens potentials, a point source near a fold or cusp caustic should yield well-defined flux ratios. For fold systems such as \sdss, the two close images of a point source should have the same brightness; deviations from this may indicate millilensing by clumped matter distributions in the lensing galaxy close to the line of sight to one or both of the images  \citep{Mao1998a, fass, Metcalf2001CompoundHalos, Metcalf2002FluxLenses,Chiba2002ProbingGalaxies, Saha2007MesoStructureSystems, Kratzer2011ANALYZINGJ1029+2623,Dalal2002}. 
 The mass scale to which these flux anomalies are sensitive can be as small as $10^6~{\rm M}_{\odot}$, sufficient to distinguish between CDM and WDM \citep{Miranda2007ConstrainingLensing, Gilman2019ProbingHaloes}. 
 At larger scales, gravitational imaging of optically-bright lens systems with extended sources has been used to more directly detect the influence of $\sim10^9{\rm M}_\odot$ substructure \citep{Vegetti2009BayesianGalaxies, Vegetti2010QuantifyingJ120602.09+514229.5, Vegetti2010DetectionImaging,Hezaveh2016DetectionSdp.81}.

Studies (eg. \citealt{xulos}) also attribute the flux anomaly to structures present along the line of sight to the primary lens  and other factors such as propagation effects \citep{Xu2015HowAnomalies}. It has been found that anomalies could also be caused by extended sources \citep{Amara2006SimulationsSubstructure, jac15}, or edge-on disc components in the lensing galaxy \citep{Moller2003DiscsH0, hsueh}. 


In addition to millilensing effects, optical sources \rev{can also be prone to producing} measurable flux variation in the lensed images due to microlensing by stars in the lensing galaxy \citep{Chang1979FluxPath}. Anomalous flux ratios due to microlensing by stars in the lensing galaxy have been found in many systems (e.g. \citealt{Irwin1989PhotometricEvent, Witt1995OnLenses, Burud2002An2149-2745, Vuissoz2007COSMOGRAIL:J1650+4251, Poindexter2008TheDisk, Vuissoz2008COSMOGRAIL:J2033-4723}); typically, monitoring of lensed quasars over the timerange of years is essential to identify the effects of microlensing.

The sample of strong lenses is currently dominated by those lenses where the source is an optically selected quasar, so only a small ($\simeq$10\%) subset of these samples have been used for substructure studies. Measurements at longer wavelengths, where the continuum source size is $\gg$~1$\upmu$as and not subject to microlensing, are important 
(\citealt{Chiba2005SubaruLensing, Fadely2011Near-infraredQuasars}, \citeyear{Fadely2012Substructure0435-1223}
;\citealt{Hezaveh2013DarkGalaxies,jac15, har19}) as are studies of the extended optical narrow emission lines \citep{Moustakas2003DetectingLenses, Sugai2007IntegralSubstructures, Nierenberg2017ProbingGrism, Nierenberg2020DoubleGrism}. 

\subsection{The quasar SDSS J0924+0219}

SDSS~J0924+0219 was discovered by \citet{Inada2003} as part of the Sloan Digital Sky Survey Quasar Lens Search (SQLS) and consists of a radio-quiet quasar at $z = 1.524$ quadruply imaged by an early-type galaxy at $z=0.394$ \citep{Eigenbrod2006COSMOGRAIL:Rings}. It was selected for observation due to an unusually large flux anomaly: it is a fold-configuration lens, but the pair of merging images, A and D in Fig.~\ref{fig1}, has a brightness ratio in the optical waveband of 12:1  \citep{Inada2003}. Spectroscopy using the Hubble Space Telescope \citep{Keeton2006} revealed the presence of flux anomaly in the broad emission line flux ratios, with the continuum and broad line emission flux ratios in the merging images being 19 and 10, respectively. This analysis also suggested the presence of microlensing in the lens system (see also \citealt{Morgan2006}). The microlensing duration expected for the lens in this system is 0.39 years and the duration averaged over a sample of 87 lenses is 7.3 months \citep{Mosquera2011TheQuasars}. However, the anomaly in SDSS J0924+0219 has persisted for 15 years now \citep{Faure2011, MacLeod2015}, which is an unusual timescale for microlensing to occur (\citealt{2004AAS...205.2806P} and \citealt{Morgan2006} predicted brightening of the demagnified image over the timescale of roughly a decade). Such an extreme flux anomaly over such a long period of time\footnote{The more recent photometry by Gaia Data Release 2 \citep{gaia} shows persistence of the flux ratio anomaly.} suggests the possibility that it could be driven by not only microlensing but also other factors such as millilensing by substructure. 



We observed \sdss\ in order to investigate the properties of the radio and molecular gas \rev{reservoirs} in the source, and investigate the causes of the flux anomaly further. This paper is divided into four sections. Section \ref{sec:obser} discusses the Karl G. Jansky Very Large Array (VLA) and Atacama Large Millimetre Array (ALMA) observations in the radio and submillimetre wavebands, and the initial data reduction carried out using the calibration pipelines of the observatories. Section \ref{analysis} discusses the data analysis in the image plane and the $u-v$ plane, followed by Section \ref{dc} which discusses the nature of the radio source and the flux anomaly in \sdss.  Where necessary we assume a flat Universe with $H_0=68.3$~km$\,$s$^{-1}\,$Mpc$^{-1}$ and $\Omega_m=0.3$; for the source this gives a luminosity distance $D_L=11400$~Mpc and an angular size distance $D_A=1790$~Mpc. \rev{1~mas corresponds to 8.6~pc at the redshift of this source}.

\section{Observations and data processing}
\label{sec:obser}


\subsection{VLA observations}


SDSS J0924+0219 was observed using the VLA in the A configuration, providing a maximum baseline of 36 km and resolution of 350 mas at 5 GHz. The observation was made under proposal code AJ413 (PI Jackson) in the C-band (4.488--6.512 GHz) with a total bandwidth of 2 GHz using 16 IFs of 128 MHz each. 12 hours of observations were divided into four blocks of three hours each, which were observed on 2018 April 4, 6, 7 and 8. \rev{The observation log is summarised in Table \ref{obslog}}.

\begin{table}
\normalsize
\caption{Summary of observations of SDSS J0924+0219 on the VLA}
\centering
\begin{tabular}{cc}
\hline
\multicolumn{1}{|c|}{\textbf{Parameter}} & \multicolumn{1}{c|}{\textbf{Values}} \\ \hline
Proposal code                   & AJ413                       \\
Observation dates               & 2018 Apr. 4, 6-8            \\
Configuration                   & A                           \\
Central Frequency                       & 5.5 GHz                      \\
Bandwidth                       & 2 GHz  
        \\
Flux Density        
            & $90\pm9.1$ $\upmu$Jy     
\end{tabular}
\label{obslog}
\end{table}

J0914+0245 and J1331+3030 (3C286) were used as the phase and the flux calibrators respectively. Each of the 3-hour blocks contained 2h 10m of on-source observation and 32~minutes on the phase calibrator; the phase calibrator was observed for 2 minutes in each 10-minute cycle, followed by 8 minutes of the target observation.

\begin{figure}

    \includegraphics[scale=0.8]{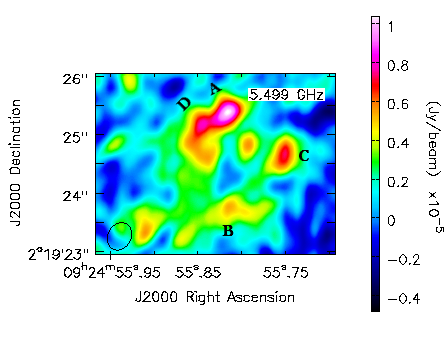}    
 
    \caption{Map of SDSS J0924+0219 from the Very Large Array. The map is the coadded version of all four 3-hours observing chunks. All four images of the lensed source are clearly detected, with a measured map noise level of 1.3~$\upmu$Jy/beam. \rev{The flux density of the lensing system in the map is $90\pm9.1$ $\upmu$Jy, assuming a 10\% flux calibration accuracy}. The beam size is 0\farcs489 $\times$ 0\farcs395 with position angle of -28.43$^{\circ}$.}
    \label{fig1}
 
\end{figure}

The VLA Calibration Pipeline, using the software package {\sc casa} (Common Astronomy Software Application\footnote{Distributed by the National Radio Astronomy Observatory; \url{http://casa.nrao.edu}.}; \citealt{Mcmullin2007CASAApplications}) was run on each of the four measurement sets. The pipeline applies basic radio frequency interference (RFI) flagging to the data, derives the initial and final delay, bandpass, and gain/phase calibrations, determines the overall flux scale of the data using the flux calibrator (J1331+3030 in this case) and also creates images of the calibrators. We combined the calibrated measurement sets generated by the pipeline and imaged the lensed source using the {\sc clean} algorithm \citep{Hogbom1974ApertureBaselines} implemented in {\sc casa} with natural weighting. The final map is shown in Fig.~\ref{fig1}. The overall off-source noise level in the map is 1.3$\upmu$Jy/beam, which is in agreement with the expected noise level of 1.1-1.3$\upmu$Jy/beam with natural weighting. \rev{The peak signal level of the map is about eight times the noise level}.


\subsection{ALMA observations}

ALMA observations \rev{(Table~\ref{obslog_alma})} were performed on 2018 December 19 (project code 2018.1.01447, PI Jackson) using \rev{45} antennas of the 12-m array, centred on the source SDSS~J0924+0219. 5 minutes of on-source integration time were obtained, with observations being performed on the nearby sources \rev{J0854+2006 and J0930+0034 for amplitude and phase calibration, respectively}. The observations were reduced by the standard ALMA pipeline system using the {\sc casa} data analysis package. Observations were performed in four spectral windows close to frequencies of 226, 228, 242 and 244~GHz; the second window was expected to contain the redshifted CO(5-4) line which has a rest frequency of 576.36~GHz and therefore an expected observed frequency of 228.35~GHz given the published redshift of 1.524. Each spectral window was divided into 128 15.6-MHz channels. The array configuration used gave a resolution of approximately 0\farcs8 with natural weighting; the calibrated data were also re-imaged with Briggs weighting (with a default robustness parameter of 0.0), giving a higher resolution of 0\farcs68  $\times$ 0\farcs56 with PA = 69.77$^{\circ}$, with only marginal increase in noise level. A VLA image was also created, for comparison purposes, with a beam convolved to match the Briggs-weighted ALMA image.

\begin{table}
\normalsize
\caption{Summary of observations of SDSS J0924+0219 on ALMA}
\centering
\begin{tabular}{cc}
\hline
\multicolumn{1}{|c|}{\textbf{Parameter}} & \multicolumn{1}{c|}{\textbf{Values}} \\ \hline
Project code                   & 2018.1.01447.S                       \\
Observation date               & 2018 Dec. 19            \\
Maximum Baseline Length                   & 500~m                           \\
Spectral Windows Frequencies                      & 226, 228, 242, 244 GHz                       \\
Total Bandwidth                       & 8 GHz  
        \\
CO(5-4) Peak Flux Density        
            & 3.9 mJy/beam \\
CO(5-4) FWHM                & $200.02\pm6.57$ km/s  \\

CO(5-4) Luminosity, $\upmu {\rm L}^{\prime}_{\rm CO}$    & $3.16\times10^{10}$\kkmspc

\\

\end{tabular}
\label{obslog_alma}
\end{table}

\begin{figure*}
    \begin{tabular}{cc}
    \includegraphics[scale=0.7]{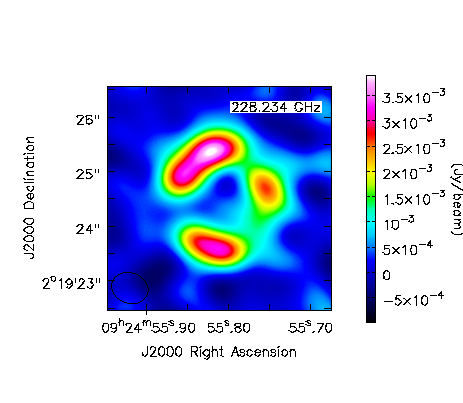}
    \includegraphics[scale=0.65]{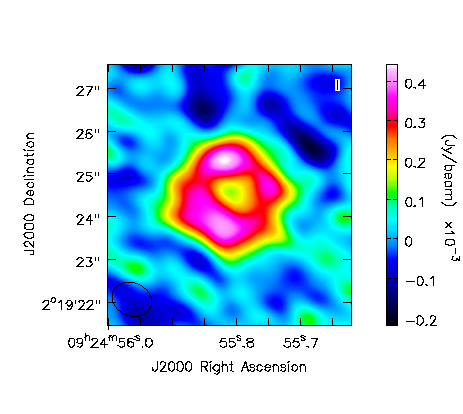}
    \end{tabular}
    \caption{Left: The CO(5-4) integrated line emission with the continuum subtracted. This image was obtained using CLEAN in CASA with Briggs weighting, giving a resolution of 0\farcs68$\times$0\farcs56 in PA 69.77$^{\circ}$. The noise level in this map is 0.2 mJy/beam. \rev{The flux density is $21.99\pm2.21$ mJy, assuming a 10\% flux calibration accuracy}. Right: Map of the submillimetre continuum obtained by cleaning the line-free spectral windows of the ALMA data with Briggs weighting. \rev{The flux density is $2.3\pm0.2$ mJy, assuming a 10\% flux calibration accuracy}.}
    
    \label{51719cont}
\end{figure*}

\begin{figure*}
\includegraphics[scale=0.6]{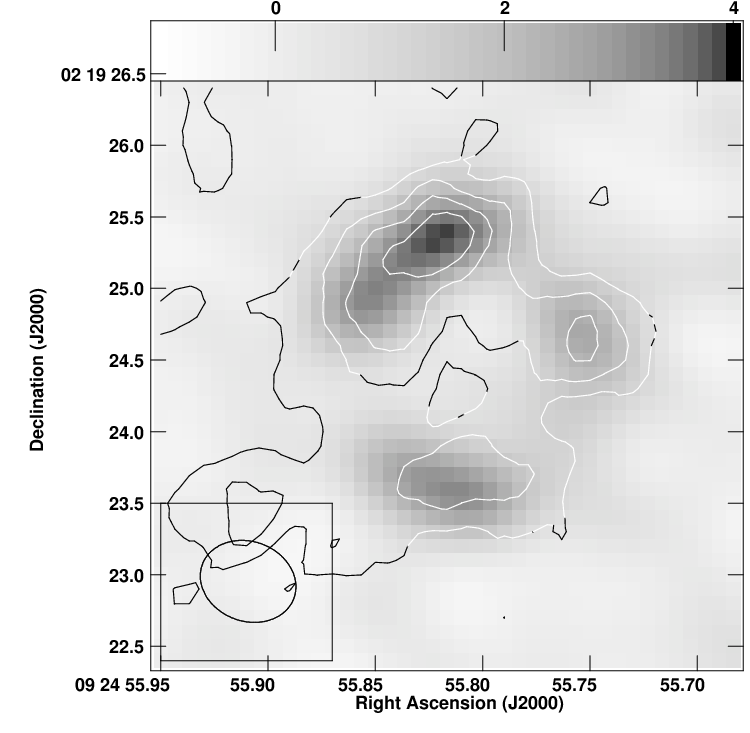}
\caption{ALMA image (greyscale) of the integrated CO(5-4) emission with VLA radio contours overplotted. The greyscale runs from \rev{$-$1.0 to 4 mJy/beam}, and the radio contours are given at \rev{(2, 4, 6, 8, 16, 32, 64, 128)} times the rms noise level. The radio image has been convolved down to the ALMA beam.}
\label{vlaalma}
\end{figure*}

\begin{figure*}
\centering
    \begin{tabular}{cc}
    \includegraphics[width=9cm]{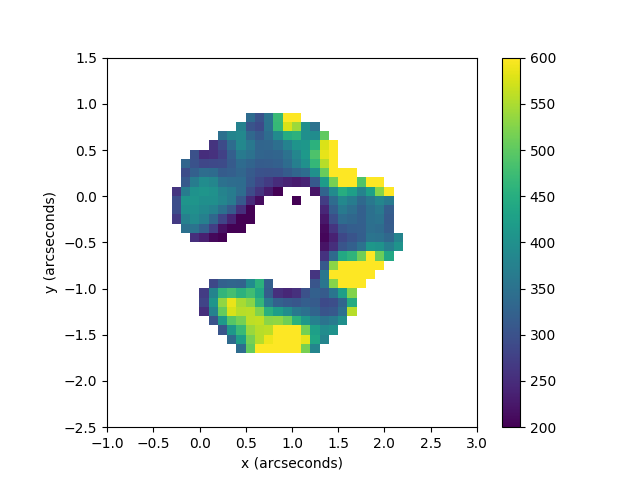}
    \includegraphics[width=9cm]{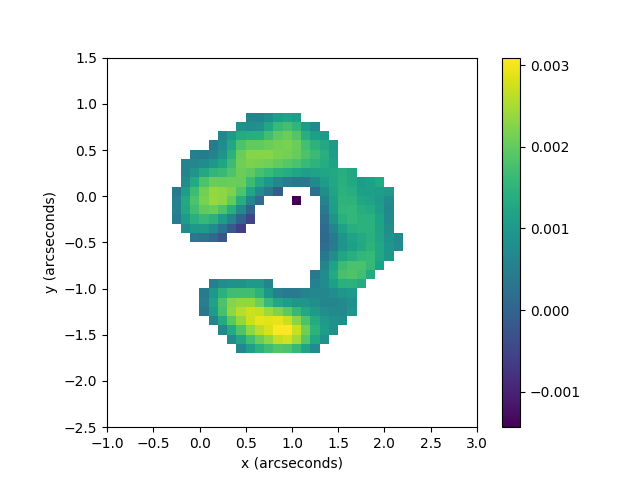}
    \end{tabular}
    \caption{Left: Map of the CO line emission divided by the VLA map convolved down to the ALMA beam. The x and y values are in arcseconds. Right: Map showing the ALMA image (normalised) subtracted by the VLA image (normalised). The ALMA and the VLA images (convolved to the ALMA beam) were, each, normalised by dividing each of them by the total flux density in the images, as the order of the flux densities in the two images are different.}
    
    \label{divmap}
    
\end{figure*}

\begin{figure}
\hskip -0.1cm    
\includegraphics[scale=0.6]{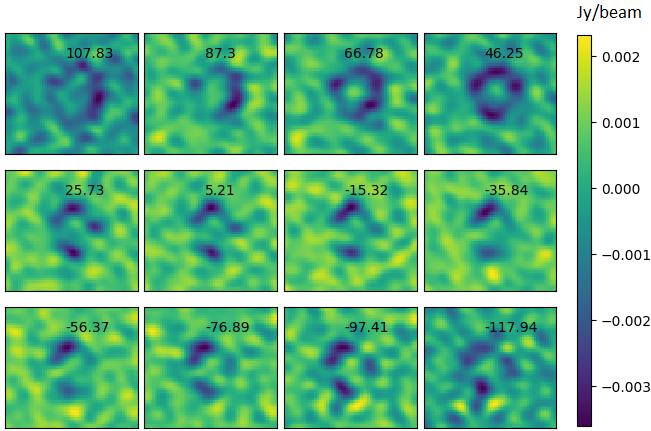}
    \caption{Slices through the ALMA CO(5-4) image cube at different frequencies; velocities on each plot are given in km$\,$s$^{-1}$ with respect to a redshift of 1.5254, the centroid of the CO emission. The colourbar is in unit of Jy/beam.}
    \label{almaco}
\end{figure}

\begin{figure*}
    \begin{tabular}{cc}
    \includegraphics[scale=0.6]{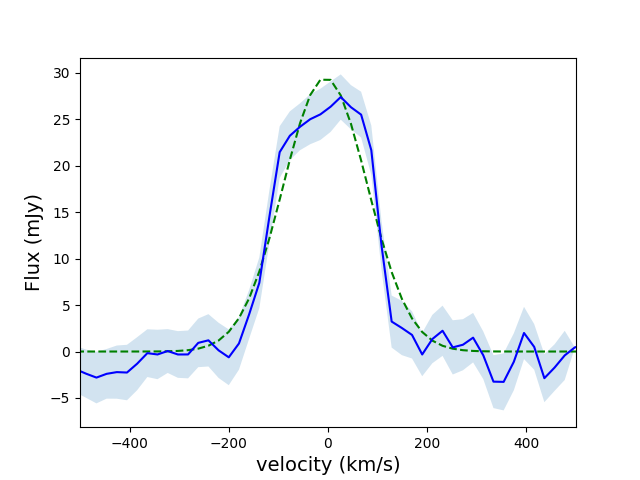}
    \end{tabular}
    \caption{
    Integrated CO(5-4) line spectrum, with the velocity centre at z=1.5254. \rev{The blue curve is the actual spectrum and the green dashed curve the gaussian fit. The shaded light blue region represents the 1$\sigma$ uncertainty on the spectrum}.}
    \label{zdet}
\end{figure*}

\begin{figure*}
    \begin{tabular}{cc}
    \includegraphics[scale=0.43]{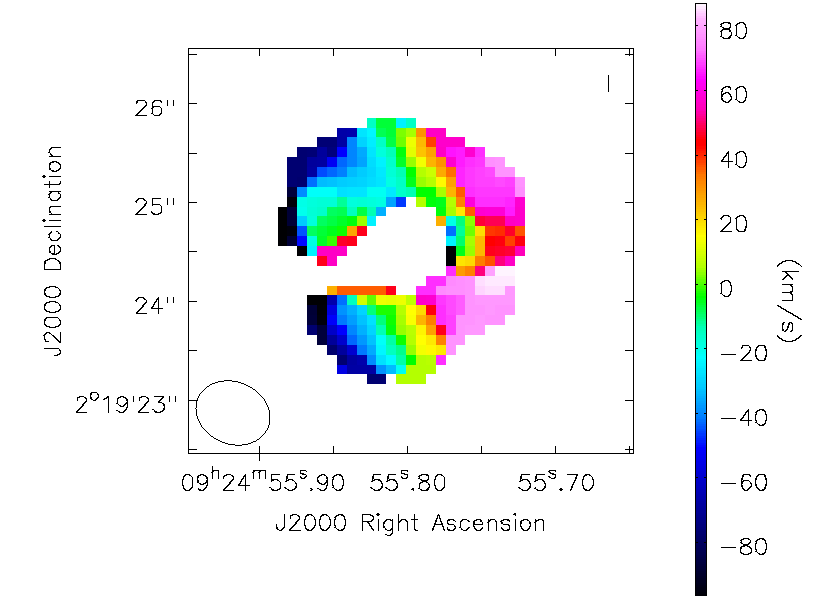}
    \includegraphics[scale=0.43]{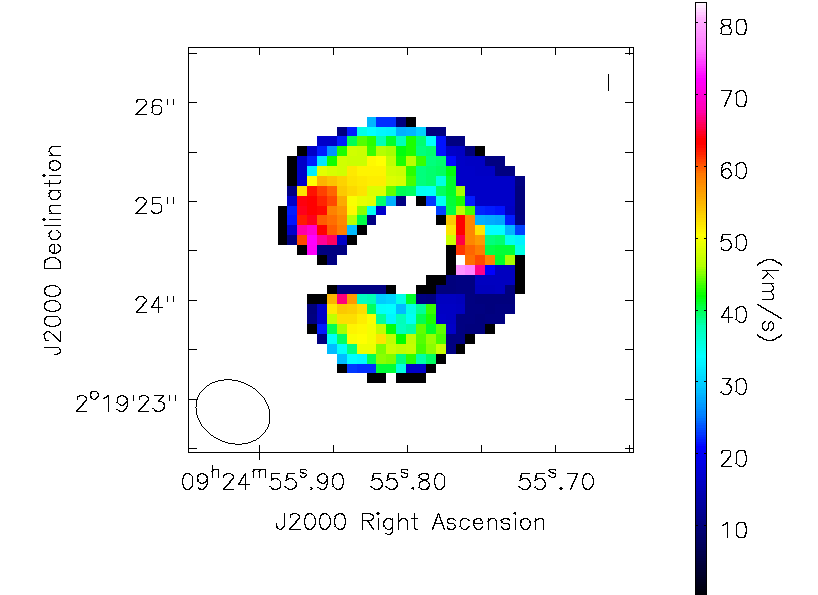}
    \end{tabular}
    \caption{Left: ALMA first-moment map showing the velocity structure across the lensed images, with the velocities being with respect to the centroid of the line emission (found to be at 228.22 GHz, thus corresponding to a redshift of z = 1.5254). \rev{Right: Moment two map (velocity dispersion) of the continuum-free CO line emission. To obtain both these maps, we included only the pixels that have a signal to noise (off-source noise in the channel maps) ratio greater than 4.}}
    
    \label{veldispersion}
\end{figure*}

In this very short observation, we achieve noise levels of 0.2~mJy/beam in each spectral window, or 0.1~mJy/beam overall. We detect the submillimetre continuum at a level of approximately 2~mJy within an area encompassing all four lensed images, but with the signal-to-noise available it is not possible to make sensible measurements of the individual component fluxes. Fig. \ref{51719cont} shows the map of the continuum obtained by cleaning the line-free spectral windows in the ALMA data. The CO(5-4) line is clearly detected in all four components, and the integrated line flux density, with the continuum subtracted, is also shown in Fig.~\ref{51719cont} (left). \rev{The peak signal to noise ratio of the CO(5-4) line map is 19.5}. Fig.~\ref{vlaalma} shows the same map in greyscale with the radio contours overlaid on top; there is a clear similarity between the two maps. A more quantitative comparison of the two maps can be seen in Fig. \ref{divmap}, which shows the CO line emission map divided by the radio map (convolved down to the ALMA beam). 
The radio and CO emission in the A and D regions appear indistinguishable, suggesting that the CO and the radio emission originate at least in part from the same region.  Velocity-resolved maps of the CO line are shown in Fig.~\ref{almaco}, and show clear velocity structure across the source, with parts of the source moving at different velocities being in different places in the source plane and thus producing different image configurations \rev{(also see Figure \ref{zdet} for the integrated CO(5-4) line spectrum)}. The first-moment map (with the continuum subtracted from the CO line emission) showing the velocity structure across the lensed source \rev{and the second-moment map showing the velocity dispersion} can be seen in Fig. \ref{veldispersion}.

\section{Analysis and results}
\label{analysis}

In the following sections we describe the analysis of the \rev{ALMA CO(5-4) line image, the submillimetre continuum, and the VLA image. Our normal procedure is to obtain an initial model in the image plane, directly by fitting to the components of the image or using a lens model to fit a lensed source. This analysis is then used as a starting point for analysis of the data in the $u-v$ (Fourier) plane. This analysis is done either by fitting the image as separate components using {\sc uvmultifit} \citep{Marti-Vidal2014UVMULTIFIT:Data} or using a lens model and the {\sc visilens} package \citep{Hezaveh2013AlmaGalaxies,Spilker2016ALMAREDSHIFTS} with a parametrised lensed source. We describe both phases of this process, firstly because the initial starting model is typically necessary to get good convergence in the $u-v$ fitting part of the process, and because it allows easy visual assessment of the fits.}

\subsection{ALMA observations: the molecular disk and submillimetre continuum}

\subsubsection{The integrated line emission}
\label{ilelm}

\begin{figure*}
\begin{tabular}{cc}
\multicolumn{2}{c}{\includegraphics[scale=1]{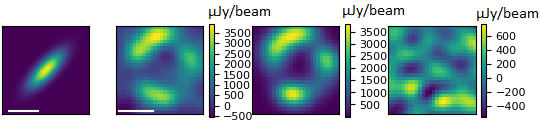}}\\
    \subfloat[\label{fig8:c}][]
    {\includegraphics[width=10cm]{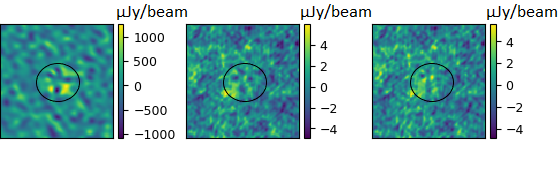}}&
    \subfloat[\label{fig8:d}][]
    {\includegraphics[width=8.5cm]{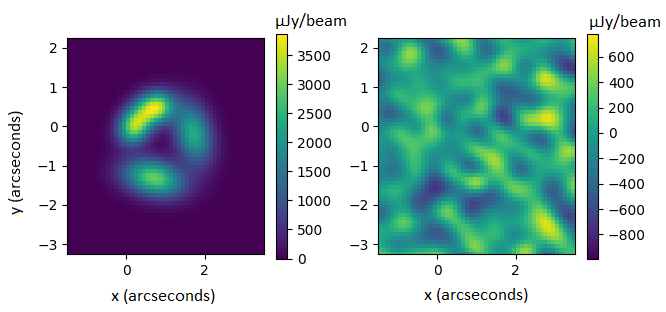}}\\
  \end{tabular}
\caption{Image-plane fitting to ALMA CO line and radio images. Top: The best fit lensed image for the SIE lens model by \citet{Eigenbrod2006COSMOGRAIL:Rings}. The first image from the left is the source structure, the second the CO line image from the data, the third the lensed image and the fourth the corresponding residual map.  The bars in the first and the second figures from the left represent 0\farcs1  in the source figure and 1\arcsec  in the lensed image, respectively. The flux density is in unit of $\upmu$Jy/beam. $\chi^2_{\rm red} = 1.85$. Bottom left ((a)): Residual maps corresponding to the best fit lens model for a point source, projecting to four PSF-sized lensed images (inside the oval region). From the left: residual for 1) CO line map 2) radio map convolved to ALMA beam and 3) radio map at its original resolution. It can be clearly seen that there is an over-subtraction of flux at the position of the images and under-subtraction in the nearby areas. The flux density units are in $\upmu$Jy/beam. Bottom right ((b)): Best fit CO-line lensed image and the corresponding residual map obtained by fitting four elliptical Gaussian components at the image positions ($\chi^2_{\rm red} = 1.5$). The A/D flux ratio for this best fit model was found to be 1.07. The unit of flux density in the image is $\upmu$Jy/beam and the x and y values are in arcseconds.}
\label{almacofit}
\end{figure*}

\rev{In the image plane, we began by fitting four elliptical Gaussians} at the positions of the images in the CO(5-4) line map, an approach that makes no assumption about the detail of the lens model, in particular, its substructure content. The fit used 10 image parameters, namely the peak flux densities of the four components, the FWHMs and the axis ratios of images A and B and two overall positional offsets in the x and y directions respectively to allow for astrometric differences in the radio and optical images. The FWHMs and axis ratios of the merging images (A and D) are assumed equal, as are the FWHMs and axis ratios for images B and C, and the fitted components were constrained to have the major axes tangential to the centre of the lensing galaxy \rev{(100.87$^{\circ}$, 77.21$^{\circ}$, 12.99$^{\circ}$ and 151.07$^{\circ}$ for A, B, C and D respectively; derived from positions of the images in \citet{Inada2003}. Since the lensed images are spatially resolved}, the FWHMs of the images were constrained to be greater than the major axis of the dirty beam. The 10-parameter optimization yielded a minimum $\chi^2_{\rm red}=1.5$. The resulting lensed image and residual for this best fit is shown in Fig. \ref{almacofit} (bottom right, Fig. (b)). Table \ref{tableimg} (values in brackets) shows the parameters of the best fit model. The A/D flux ratio for this model was found to be 1.07. The modelled images B and C are clearly wider than would be expected in any plausible gravitational lens model.

\rev{In addition, we performed a gravitational lens modelling\footnote{Throughout this study, we have used the lens modelling code developed by Neal Jackson.} using the Singular Isothermal Ellipsoid (SIE) model developed by \citet{Eigenbrod2006COSMOGRAIL:Rings} for the lensing galaxy}. Allowing the source to have a Gaussian profile, the galaxy position and the source parameters were optimized using the difference between the data and the lensed images convolved with the point spread function. An extra parameter, $f_{\rm AD}$, was introduced, that can vary between 0 and 1 and which multiplies the flux of the component D, while keeping the flux of A the same. This parameter quantifies the anomaly in the flux ratio of A and D due to substructure effects, with a value of 1 indicating the situation of no flux anomaly present due to CDM substructure. \rev{Due to the relatively low signal to noise ratio of the data, we did not attempt to pixelize the source and use semi-linear source reconstruction \citep{war03}}. The best-fit lensed image, source structure and the corresponding residual map obtained from the optimization are shown in Figure \ref{almacofit} (top). \rev{The fit has a $\chi^2_{\rm red}=1.85$ and is generally satisfactory apart from some difference in the structure of image B}. The parameters for this model are shown in Table \ref{tablesrc} (values in brackets). \rev{The best fit position of the lensing galaxy from the optimization corresponds to RA = 09h\,24m\,55.813s and Dec = 02$^{\circ}$\,19$^{\prime}$\,24\farcs454 and has been used for all subsequent image-plane analysis, wherever required}. The value of $f_{\rm AD}$ for the best fit is 1. We also attempted to fit a lens model assuming a point source, giving unresolved point images\footnote{The fit was performed using {\sc jmfit}, part of the Astronomical Image Processing System, {\sc aips}, which is distributed by the US National Radio observatory, \url {http://www.aips.nrao.edu}.}; the resultant residual shows an over-subtraction of flux in the image positions and under-subtraction in the nearby areas (see Fig. \ref{almacofit}, bottom left), demonstrating incompatibility of a point source with our observation. The observation that $f_{\rm AD}=1$ implies that the SIE model we used clearly reproduces the main features of the data and is compatible with an extended source, together with no requirement for additional perturbations due to millilensing by substructure. 

\begin{figure*}
    \centering
    \includegraphics[width=16cm]{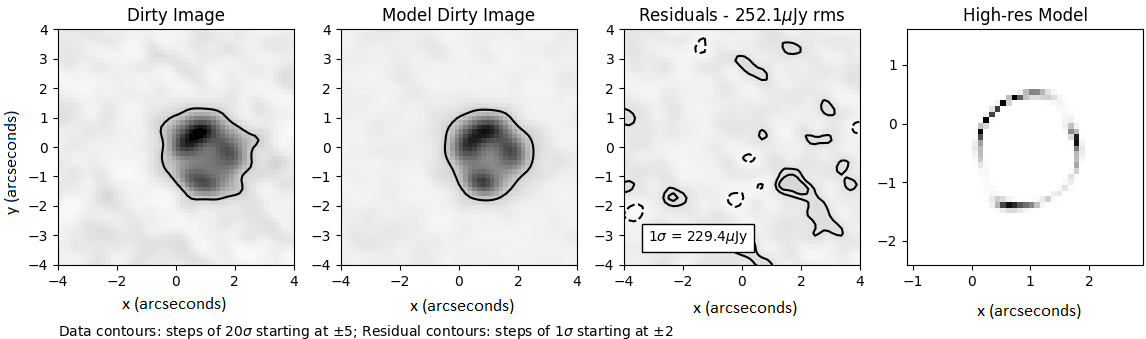}
    \caption{Best fit model obtained by fitting the \citet{Eigenbrod2006COSMOGRAIL:Rings} SIE model to the visibility data of the CO(5-4) emission line, obtained using {\sc visilens}.}
    \label{almacofit_uv}
\end{figure*}

\begin{figure*}
 \includegraphics[width=19cm]{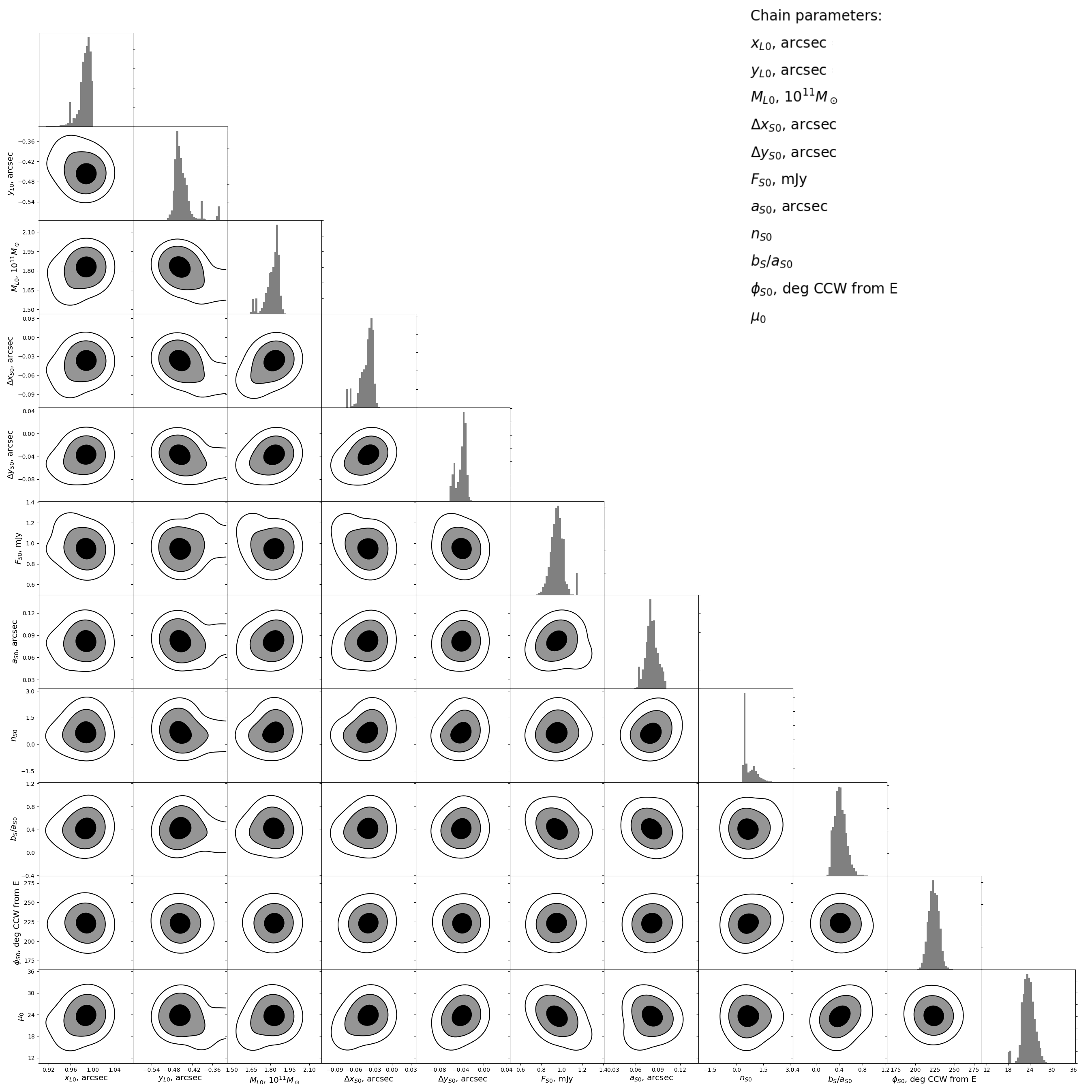}
    \caption{Corner plot showing the distribution for different parameters, obtained by fitting a lens model to the CO(5-4) visibility data using {\sc visilens}, described in Section \ref{ilelm}. For this optimisation, all lens galaxy parameters were kept fixed, except the lens galaxy mass and position. We also optimised for all source parameters. The parameters (with their units) are shown in the upper right corner for clarity. They refer to: position of lensing galaxy, mass of the lensing galaxy, position of source with respect to the lensing galaxy, flux of the CO source, width of the source, Sersic index, source axis ratio, phase angle and magnification.}
    \label{CO_UV_cornerplt}
\end{figure*}

\rev{Given a starting model derived from the image plane, we now derived quantities directly from the $u-v$ data. We first fitted four elliptical gaussians for the integrated CO line using {\sc uvmultifit} \citep{Marti-Vidal2014UVMULTIFIT:Data}, on the continuum subtracted visibility data of the CO line. We used the positions of the four images from \citet{Inada2003}  and kept them fixed, optimising for the flux densities, widths and axes ratios of the images. We kept the axes ratios of A and D (and likewise B and C) equal, as well as their FWHMs. The parameters corresponding to the best fit from the optimization can be found in Table \ref{tableimg}. The flux ratio A/D obtained from this fitting was found to be $1.58\pm0.15$, which is slightly bigger than the corresponding flux ratio from the image plane analysis.

We also fitted the $u-v$ data with a lens model using {\sc visilens}. All the lens galaxy parameters, except the position and mass of the lens galaxy, were fixed assuming the SIE lens model by \citet{Eigenbrod2006COSMOGRAIL:Rings}. We optimised for all the source parameters and the lens galaxy mass and position. Figure \ref{almacofit_uv} shows the best fit model and the residual map. The parameters for the best fit can be found in Table \ref{tablesrc}. The source FWHM using the $u-v$ plane analysis is 0\farcs081$\pm$0\farcs008, slightly larger than the FWHM found in the image plane analysis. Figure \ref{CO_UV_cornerplt} shows the distribution of the different parameters resulting from this lens model fitting.} 

We now proceed to calculate the luminosity of the CO(5-4) emission. Following \citet{sol05}, we get a CO line (lensed) luminosity of \rev{$\upmu {\rm L}^{\prime}_{\rm CO}=3.16\times10^{10}$\kkmspc. The lens model fitting of the $u-v$ data using {\sc visilens} demonstrates a source magnification of 23.5, making the intrinsic CO line luminosity equal to ${ \rm L}^{\prime}_{\rm CO}\simeq 1.34 \times 10^{9}$\kkmspc. \citet{sol05} use the formula ${\rm M}_{\mathrm{dyn}}$sin$^{2}(\mathrm{i}) = 233.5 {\rm R} \Delta $V$^{2}$ to calculate the dynamical mass, where R is the radius of the molecular disk and $\Delta$V is the FWHM of the CO line. Using our value of size of the integrated CO line emission source of $81\pm8$ mas (Table \ref{tablesrc}), we get R = $348\pm34$ pc, approximating R to be equal to half of size of the disk. The FWHM of the line is $200\pm7$ km/s, thus giving a dynamical mass ${\rm M}_{\mathrm{dyn}} =  3.25\times10^{9}$cosec$^{2}(\mathrm{i}){\rm M}_{\odot}$ where i is the inclination angle of the disk}.

\begin{table*}
\normalsize
\caption{Parameters corresponding to the best fit model to image components obtained by fitting elliptical gaussians to the observed lensed images, in the integrated CO(5-4) line and the radio. Refer to the text for the priors on the model. The reduced chi squared value before the re-scaling of the visibility weights in the algorithm are 4.68 and 11.24 for the CO line and the radio, respectively. The flux densities for the CO line and the radio data are in mJy and $\upmu$Jy, respectively. FWHMs are in arcseconds. Quantities derived from the initial image-plane fitting are given in brackets, in the case of the radio data for the radio image convolved to the ALMA resolution; other data are from {\sc uvmultifit} fitting to the $u-v$ data. The bottom table shows the positions of the four images with respect to the lensing galaxy. For the $u-v$ fit, the positions from \citet{Inada2003} were used and kept fixed. Positions in square brackets correspond to the best fit model in the image plane fits.}
\begin{supertabular}{lllllll}
\cline{2-7}
\multicolumn{1}{l|}{} & \multicolumn{2}{c|}{\textbf{Flux Density}}                         & \multicolumn{2}{c|}{\textbf{FWHM}}                                 & \multicolumn{2}{c|}{\textbf{Axis ratio}}                           \\ \cline{2-7} 
\multicolumn{1}{l|}{} & \multicolumn{1}{l|}{CO line} & \multicolumn{1}{l|}{Radio} & \multicolumn{1}{l|}{CO line} & \multicolumn{1}{l|}{Radio} & \multicolumn{1}{l|}{CO line} & \multicolumn{1}{l|}{Radio} \\ \cline{2-7} 
\textbf{A}                     & $5.39\pm0.30$ (4.5)                & $16.71\pm3.41$  (16.8)           & $0.48\pm0.07$ (0.8)                & $0.32\pm0.12$  (0.80)            & $0.75\pm0.19$ (0.83)               & $0.69\pm0.26$   (0.9)           \\
\textbf{B}                     & $6.71\pm0.36$ (7.1)                & $12.18\pm3.16$  (13.2)           & $1.32\pm0.07$ (1.25)               & $0.32\pm0.12$  (0.79)            & $0.35\pm0.04$ (0.58)               & $0.69\pm0.26$   (1.0)           \\
\textbf{C}                     & $7.41\pm0.41$ (5.7)                & $13.55\pm3.14$  (14.1)           & $1.32\pm0.07$ (1.25)               & $0.32\pm0.12$  (0.80)            & $0.35\pm0.04$ (0.58)               & $0.69\pm0.26$   (1.0)           \\
\textbf{D}                     & $3.41\pm0.26$ (4.2)               & $13.06\pm3.25$   (13.5)          & $0.48\pm0.07$ (0.8)               & $0.32\pm0.12$    (0.79)          & $0.75\pm0.19$  (0.83)              & $0.69\pm0.26$   (0.9)  \\         
\end{supertabular}


\begin{tabular}{ccc}
\cline{2-3}
\multicolumn{1}{l|}{\multirow{2}{*}{}} & \multicolumn{2}{c|}{\textbf{Relative Positions (to lensing galaxy) (x, y)/\arcsec}}   \\ \cline{2-3} 
\multicolumn{1}{l|}{}                  & \multicolumn{1}{c|}{CO line}          & \multicolumn{1}{c|}{Radio}            \\ \cline{2-3} 
\textbf{A}                             & (-0.162, 0.847) {[}(-0.06, 0.8){]}    & (-0.162, 0.847) {[}(-0.09, 0.77){]}   \\
\textbf{B}                             & (-0.213, -0.944) {[}(-0.11, -0.98){]} & (-0.213, -0.944) {[}(-0.15, -1.02){]} \\
\textbf{C}                             & (0.789, 0.182) {[}(0.89, 0.14){]}     & (0.789, 0.182) {[}(0.85, 0.1){]}      \\
\textbf{D}                             & (-0.702,  0.388) {[}(-0.60, 0.35){]}  & (-0.702, 0.388) {[}(-0.63, 0.30){]}  
\end{tabular}

\label{tableimg}
\end{table*}

\begin{table*}
\normalsize
\caption{Parameters of source structures inferred from fits assuming a lens model, corresponding to the best fit for the integrated CO line, submm continuum and radio map (convolved to ALMA resolution) of SDSS J0924+0219. These were fitted in the $u-v$ plane with a S\'ersic profile source using {\sc visilens}, and (figures in brackets) with a Gaussian profile source in the image plane. $\Delta x_{\mathrm{S0}}$ and $\Delta y_{\mathrm{S0}}$ are the positions of the source with respect to the lens, in arcseconds. All angles are east of north. A standard galaxy model (critical radius 0\farcs87, ellipticity 0.13 at PA $-73.1^{\circ}$, shear magnitude 0.042 at PA 65.4$^{\circ}$) is used for the image plane fitting. For the $u-v$ plane fitting, the lens parameters are the same as for the image plane, except for the Einstein radius (0\farcs92) which was obtained by optimising for it in case of the CO line visibility data and subsequently kept fixed for the submillimeter continuum and radio $u-v$ fits. The image plane fit for the submillimeter continuum did not involve the $f_{\rm AD}$ parameter.}
\centering
\begin{tabular}{lccc}
\hline
\multicolumn{1}{|c|}{\textbf{Parameters}} & \multicolumn{1}{c|}{\textbf{CO line}} & \multicolumn{1}{c|}{\textbf{submm continuum}}&\multicolumn{1}{c|}{\textbf{Radio}}\\ \hline
$f_{\rm AD}$                  & 1                  & NA &0.97\\
$\Delta x_{\mathrm{S0}}$/\arcsec &$-0.037\pm 0.009$ ($-$0.020) &$-0.050\pm0.019$ ($-$0.026)&$-0.037\pm0.010$ ($-0.025$)\\
$\Delta y_{\mathrm{S0}}$/\arcsec &$-0.036\pm 0.010$ ($-$0.022)&$-0.017\pm0.026$ (0.005) &$-0.044\pm0.009$ ($-0.019$)\\
Source Flux Density/mJy           &$0.954\pm 0.057$ (0.8)        &$0.225\pm0.048$ (0.1) &$0.003\pm0.002$ (0.0023)\\
Source FWHM/\arcsec           &$0.081\pm 0.008$ (0.045)    &$0.260\pm0.071$ (0.21) &$0.120\pm0.089$ (0.052)\\
Source axis ratio             &$0.41\pm 0.10$ (0.36)       &$0.866\pm0.122$ (0.92) &$0.79\pm0.17$ (0.43)\\
Source PA/$^{\circ}$           &$223.85\pm 7.55$ (136.0)          &$171.50\pm131.07$ (0.0) &$162.16\pm117.79$ (125)\\
S\'ersic index                  &$0.5\pm 0.3$                &$1.029\pm0.489$&$1.784\pm0.324$\\
\end{tabular}
\label{tablesrc}
\end{table*}

\subsubsection{ALMA observations: velocity-resolved structure in the CO(5-4) line}

The ALMA CO(5-4) line observations (Fig.~\ref{almaco}) clearly show that the line originates in a structure with ordered spatial differentiation in velocity, such as rotation or possibly outflow. We can use a simple lens model to project the frequency planes of the ALMA data back into the source plane and thus determine simple parameters, such as size and rotation velocity of the emitting structure. The relatively low resolution of the ALMA observations do not allow us to ask more than simple questions of the data. In particular, we cannot derive the structure of the source in detail, nor do we have the sensitivity at high resolution needed for investigation of detailed structure within the mass model. Observations of another CO line which do allow this are presented in Stacey et al. (in preparation).

The relatively low resolution, and the complexity of the information about the source that we seek, motivate a simple analysis in which we project each of the frequency-sliced images back to the source plane, without modelling the $u-v$ data explicitly. This was the approach taken using simple VLA information on the source HS~0810+2554 \citep{jac15} which proved to yield roughly correct values for the source size when higher quality observations became available \citep{har19}. We use the previous SIE lens model (Section \ref{ilelm}) which allows us to fix the galaxy mass parameters, and we represent the source structure in each frequency plane as a single elliptical Gaussian, whose position, flux, width, axis ratio and position angle we allow to vary in order to fit the data. The resulting source positions occupy a roughly linear configuration, with a clear sequence from red to blue across the frequency band. These assumptions about the source effectively constitute a regularisation, which prevents the source structure from varying catastrophically as the model attempts to fit noise in the image.

\begin{figure*}
    \centering
    \includegraphics[scale=0.85]{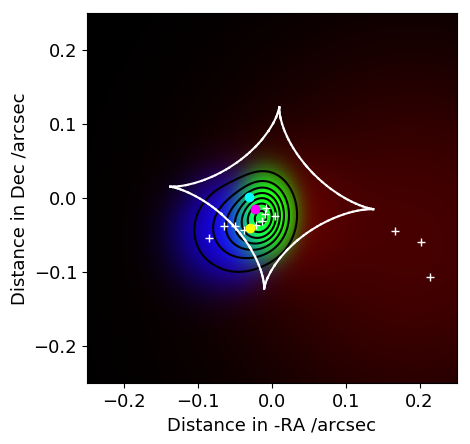}
    \caption{Reconstructed source based on the ALMA observations, coloured according to frequency of the source (see text). The astroid caustic of the model galaxy is overlaid. The white crosses are the positions of the sources for the individual CO planes. The black contours are the sum of contours of the sources in all the planes. The circles in yellow and magenta correspond to the position of the optical and the radio source, respectively. The circle in cyan corresponds to the  submillimeter continuum source (refer to Fig.~\ref{submmfit}).} 
    \label{almasource}
\end{figure*}

The resulting source model, with the astroid caustic of the galaxy overlaid on top, is shown in Fig.~\ref{almasource}. This caustic separates regions in which the source is doubly and quadruply imaged, and is a locus of \rev{infinite} magnification. The source model in Fig.~\ref{almasource} is compatible with the images of the different CO planes in Fig.~\ref{almaco}, and its position angle and ellipticity are compatible with the source structure inferred by modelling the integrated CO image. However, the interpretation of details in Figure \ref{almasource} has to be done with caution: while we may be tempted to interpret the source to be a warped disk, it should be noted that there could be a large error bar on the fit to the velocity planes towards the red end because of their low signal to noise, as can be seen from Figure \ref{almaco}. \rev{Given the limited angular resolution, the structure could be consistent with a rotating molecular disk (eg \citealt{Smit2018Rotation6.8}), which has a size of between $0\farcs1$ and $0\farcs3$ depending on the exact centroids inferred for the lower-frequency parts of the source, corresponding to $\sim850$--2500~pc at the redshift of this quasar. However, observations at higher angular resolution and sensitivity are required to confirm this hypothesis}. The size of the structure, and the overall distribution of velocities, is robust to different assumptions about the source structure, such as allowing the source components corresponding to each frequency plane to be circular Gaussians.

\subsubsection{The submillimetre continuum}
\label{sub_cont_imageplane}

\begin{figure*}
    \caption{Top, left to right: The best fit source structure, data, lensed image configuration and residual map obtained from the image-plane lens modelling of the submillimetre continuum using the SIE galaxy lens model \citep{Eigenbrod2006COSMOGRAIL:Rings} used in Section \ref{ilelm} ($\chi^2_{\rm red} = 1.19$). The bars in the first and the second figures from the left represent 0\farcs1  in the source figure and 1\arcsec  in the lensed image, respectively. The flux density is in units of $\upmu$Jy/beam. The curves in the source and the model panels show the astroid caustic and the critical curve. Bottom: Fit to the $u-v$ data of the submm continuum, starting with the image-plane fit.}
    \begin{tabular}{c}
     \hskip -10mm \includegraphics[width=15cm]{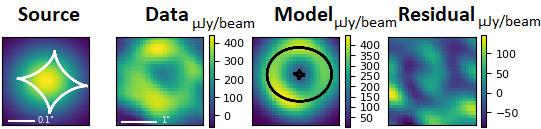}\\
\hskip -15mm      \includegraphics[width=17cm]{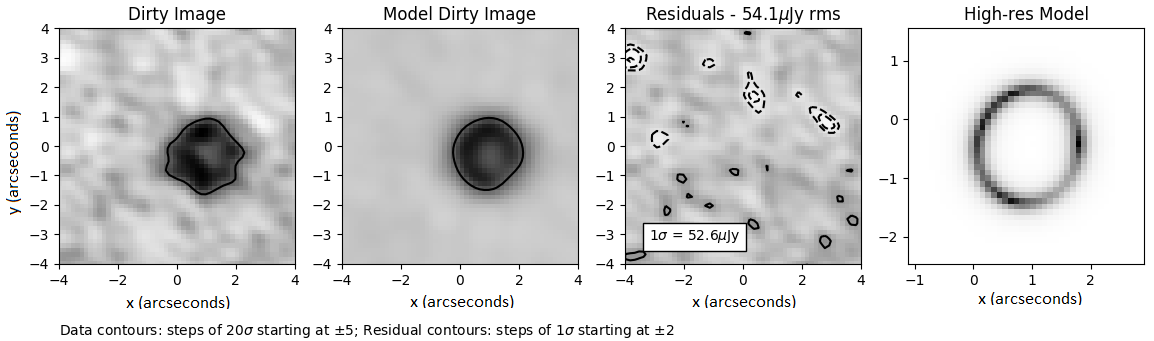}\\
      \end{tabular}
    \label{submmfit}
\end{figure*}

Fig.~\ref{51719cont} shows that the lensed submillimetre continuum displays a ring structure, but which appears subtly different in the distribution of bright features to the CO line image. A simple lens model of this structure, using the same SIE galaxy lens model as before, gives a best fit in which the continuum source is somewhat offset from the line; the best-fit source position is indicated in Fig.~\ref{almasource}. It appears closer to the northeastern fold caustic, in agreement with the appearance of the lensed continuum map which suggests a different position for the merging image in the continuum and line maps. This result requires confirmation, however, with a longer observation. 
The source structure, lensed image configuration and the residual map obtained using the SIE galaxy lens model are shown in Fig. \ref{submmfit}; the width of the continuum source from the model is about 0\farcs21 ($\approx$1.8 kpc at the redshift of this source). This is found to be more compact compared to the dust emission from a sample of dusty star-forming galaxies studied by \citet{Hodge2016Kiloparsec-ScaleGalaxies, Hodge2019ALMAGalaxies}.

\rev{The source parameters from this image-plane fit were used as initial parameters to perform a $u-v$ plane fitting using {\sc visilens}, incorporating the SIE model of \citet{Eigenbrod2006COSMOGRAIL:Rings}. For the optimization, we kept the parameters of the lensing galaxy fixed (using the lensing galaxy position and mass derived from the optimisation of the CO line visibility data using {\sc visilens} previously) and let the source parameters vary. We assumed the source to have a S\'ersic profile (Fig.~\ref{submmfit}). The size of the submillimetre continuum found using $u-v$ plane fitting (0\farcs26$\pm$0\farcs07, Table~\ref{tablesrc}) is consistent with the size of 0\farcs21 found using the image plane fitting.}

\subsection{Radio}
\label{radio_lmg}

In the VLA map of SDSS~J0924+0219 (Fig. \ref{fig1}) we detect all four lensed images in the system, and it is clear that the ratio of the fluxes in the A and D images is less than the value of 12 which is measured in the optical waveband, but that the flux densities of the two images do not look equal. \rev{We again began by fitting Gaussian components in the image plane}. A preliminary fit of the four lensed images with point-like gaussian components, using {\sc aips jmfit}, gave obviously unsatisfactory results, similar to what is seen for the CO line map (Fig. \ref{almacofit}). This shows that the radio source is also extended.



\begin{figure*}
  \caption{Top: The best fit image-plane lens model for the radio map (convolved to the ALMA beam) assuming the SIE lens model by \citet{Eigenbrod2006COSMOGRAIL:Rings} ($\chi^2_{\rm red} = 2.05$). The bars in the first and second figures from the left represent 0\farcs1 in the source figure and 1\arcsec in the lensed image, respectively. The flux density is in units of $\upmu$Jy/beam. The curves in the source and the model panels show the astroid caustic and the critical curve. Middle: (Left) Image plane fit with elliptical Gaussian components, with residual: (Right) the same, but enforcing a 12:1 ratio between components A and D. Bottom: Image, model, residual and high-resolution model from the $u-v$ plane fit.}
  \begin{tabular}{cc}
\multicolumn{2}{c}{\includegraphics[scale=1.2]{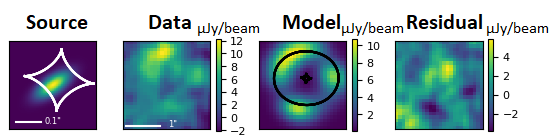}}\\
\hskip -80mm
\includegraphics[scale=0.6]{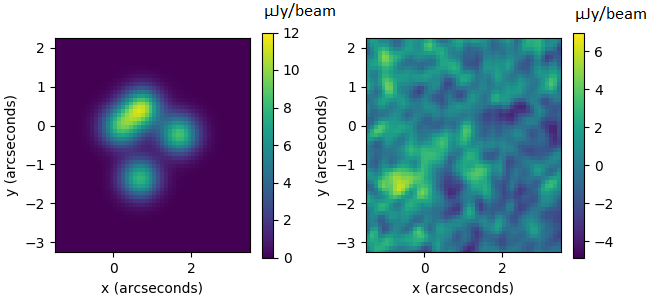}&
\mbox{}\hskip -80mm \includegraphics[scale=0.6]{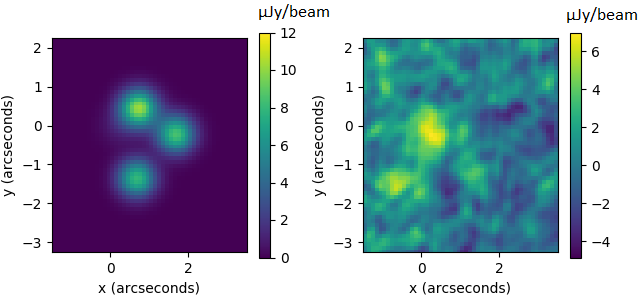}\\
\includegraphics[width=17cm]{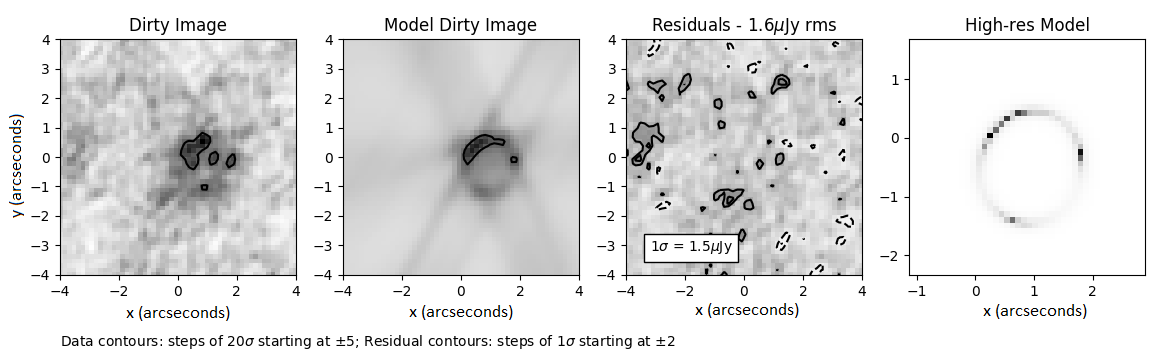}\\
\end{tabular}
\label{radiofit}
\end{figure*}

\rev{The flux ratio of components A and D was determined using the radio map, convolved to the ALMA beam size, using the Gaussian fitting method described earlier for the CO line, with the same 10 image parameters}. The best fit, of $\chi^2_{\rm red}=1.8$, yielded an A/D flux ratio of 1.25; \rev{on the original map we obtained 1.28}. The best fit model and its parameters are shown in Fig. \ref{radiofit} and Table \ref{tableimg} (values in brackets). For illustration, Fig. \ref{radiofit} shows what the lensed image and the corresponding residual would look like if we constrained the A/D flux ratio for this lens model to be 12, the ratio observed in the optical waveband. It confirms our previous result that the radio data does not demonstrate an A/D flux ratio of 12 and is more consistent with a lower flux ratio. \rev{We carried out the lens modelling of the radio image, convolved to the ALMA beam and incorporating the same SIE model for the lensing galaxy as we did for the ALMA data (Section \ref{ilelm}) and assuming a Gaussian function for the source structure}. We kept the lensing galaxy parameters \rev{fixed} and optimized only for the source and the $f_{\rm AD}$ parameter. The source structure, the best fit lensed image and the residual map corresponding to the lens model ($\chi^2_{\rm red} = 2.05$) are shown in Figure \ref{radiofit} and the corresponding parameters in Table \ref{tablesrc} (values in brackets). Similar to the CO line image, the $f_{\rm AD}$ parameter in this case is also close to 1, implying that any flux anomaly in the radio can be attributed to the extended source structure. 


\rev{These image-plane fits were used to start the fitting in the $u-v$ plane. As for the integrated CO line, we fitted four elliptical Gaussians to the visibility data using {\sc uvmultifit}, with image positions from \citet{Inada2003} and fixing the component position angles. The four image flux densities, FWHMs and axis ratios were optimized, keeping the FWHMs and axis ratios of the four images equal, because using any more parameters than these six (four flux densities, one width and one axis ratio) resulted in the program failing to find a fit. The parameters for the best fit model are tabulated in Table \ref{tableimg}. The flux ratio A/D obtained from this fitting was found to be $1.28\pm0.41$, once again consistent with the flux ratio found from the image-plane analysis.  

Following the same approach as for the CO(5-4) line, we fixed the lens galaxy parameters and optimised for the source parameters in {\sc visilens}. Figure \ref{radiofit} shows the best fit model and the residual map and Table \ref{tablesrc} shows the corresponding parameters. Even though the median size of the radio source inferred from the $u-v$ fit is seen to be only 1.3$\upsigma$ away from zero, it can be seen from Figure \ref{almacofit} that the radio data is inconsistent with a point source. We also find that the radio source size is sensitive to the starting parameters of the $u-v$ fit optimisation.}

\section{Discussion and Conclusions}
\label{dc}

There are two interesting questions we can attempt to answer with these data, the answer to the second of which is related to the first:

\begin{itemize}
    \item What is the size of the radio-emitting region, what physical processes produce it, and is it co-spatial with the molecular line gas?
    \item Is the optical flux anomaly due to microlensing?
\end{itemize}

\rev{We discuss these two points in the following sections.}

\subsection{The radio source and the CO(5-4) line}

\begin{figure}
\centering
    \includegraphics[scale=0.6]{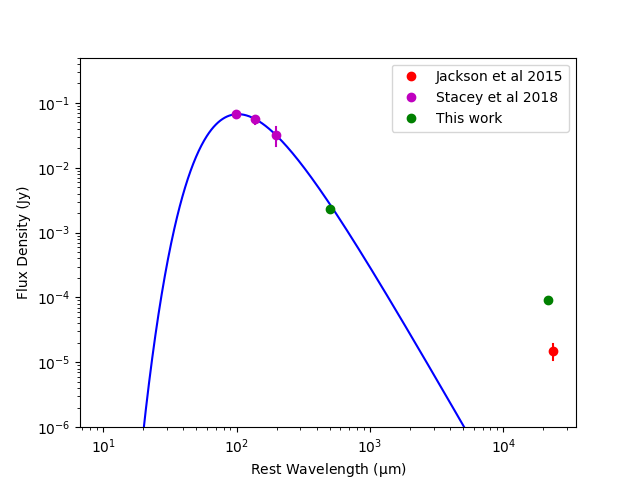}
    \caption{The spectral energy distribution of SDSS J0924+0219, corresponding to a dust temperature of $T = 30.8\pm0.8$ K and $\upbeta$=$1.6\pm0.08$. The SED was fit to the modified blackbody curve defined by $S_{\nu}$ $\propto$ $\frac{\upnu^{3+\beta}}{e^{\frac{h\upnu}{kT}}-1}$, where $h$ is the Planck constant, $k$ the Boltzmann constant, $\upbeta$ the emissivity index and $T$ the dust temperature.}
    \label{finalsed}
\end{figure}

Although we do not have the resolution to study the structure of the molecular disk in detail, \rev{it appears from the ALMA CO(5-4) observation that we could be seeing a structure in ordered motion}, the natural explanation of which is the presence of a rotating molecular disk. The structure is of physical dimension approximately \rev{100--300~mas ($\sim850$--2500~pc)}, consistent with a compact star-forming disk in the centre of the host galaxy of SDSS~0924+0219. We have performed simple modelling of the radio data (which is all that is warranted given the signal-to-noise and resolution of the data) and derive inconsistency with a point source, and a preferred size of about \rev{100~mas}, consistent with the size of the CO-line source. From Fig.~\ref{vlaalma} we see that the lensed structures in the radio and CO lines are consistent with each other. 
Although not conclusive, the most natural explanation is therefore that the two components of the source originate from a region which is of the same size and physical origin. \rev{Combining the radio flux density from this work with the previously determined far-IR flux densities of \citet{sta18}, it appears that \sdss\ lies on the standard radio-FIR correlation \citep{Ivison2010TheHerschel}; the corresponding $\rm q_{\rm IR}$ parameter is found to be 2.35, well within the range of $\rm q_{\rm IR}$ = $2.40\pm0.24$ found by \citet{Ivison2010TheHerschel}}. Thus the conclusion of radio emission produced in a star-forming region appears natural. \rev{The physical size of the disk, of order of a kiloparsec, is similar to, or slightly smaller than that found in other high redshift lensed quasars with mid-J CO transition \citep{Alloin1997TheEmission,Downes1999Detection3.9,ven03,Anh2013ResolvingRXJ0911.4+0551,ryb15,gea18}. It is also found to be similar or more compact in size to mid-J transition CO emission regions in high redshift, unlensed quasars \citep{Beelen2004StarburstJ1409+5628,Riechers2009ImagingReionization,Polletta2011DiskZ3.4,Feruglio2018TheALMA,Brusa2018MolecularALMA}}. The unlensed flux density of the CO line, as well as the molecular gas mass, are both about a factor of 10 smaller than those of the radio-loud quasars JVAS~B1938+666 and MG~0751+2716 \citep{spi19}, although the observations are not strictly comparable as the higher-order CO line in this investigation preferentially traces warmer gas, \rev{which tends to be more compact than the cold molecular gas. 
Figure \ref{finalsed} shows the spectral energy distribution (SED) of SDSS J0924+0219, based on this work and previous studies carried out by \citet{sta18} and \citet{jac15}. The SED
was found to correspond to a dust temperature of $T$ = $30.8\pm0.8$ K and emissivity index $\upbeta=1.6\pm0.08$.}

Our ALMA observations detect the submillimetre continuum source, although at a relatively low signal-to-noise level. Modelling of these observations gives a preliminary indication that the submillimetre continuum source, which traces the illuminated dust, may be spatially displaced from the region emitting the molecular line and radio emission, although this requires confirmation from \rev{deeper} observations.   Whether the radio emission contains partial contribution from an AGN-type source, as in HS~0810+2554 \citep{har19} will be interesting to determine from future VLBI observations. The submillimeter source is also more extended than the CO emitting region; it may be that star formation is taking place in only part of the overall structure of dust and gas at the centre of the object.

\subsection{Is microlensing the reason behind the optical flux ratio anomaly?}

\begin{figure}
    \includegraphics[scale=0.61]{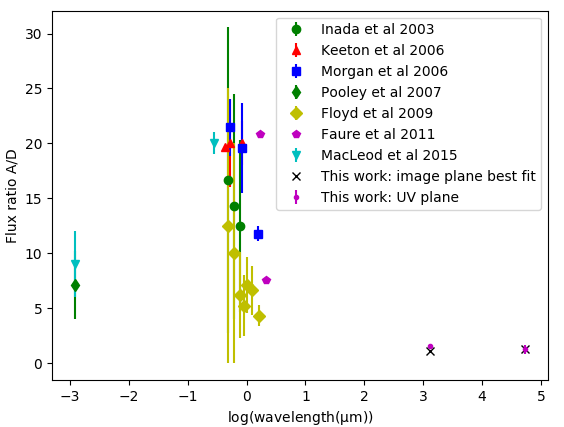}
    \caption{Flux ratios in SDSS J0924+0219 over different wavelengths. The data for this plot has been taken from Table 2 in \citet{Floyd2009TheJ0924+0219}, \citet{Keeton2006}, \citet{Morgan2006}, \citet{Faure2011} and  \citet{MacLeod2015}. The logarithm in the x-axis is in base 10.} 
    \label{frvswl}
\end{figure}

We now turn to the second question concerning the optical flux ratio anomaly. In the case of SDSS J0924+0219, the flux ratio between the brightest image and its merging counterpart in the radio, was robustly determined as \rev{1.28$\pm$0.41, consistent with all methods of analysis}. Similar results are obtained from the CO line images, and at this resolution the observations are consistent with a combination of a smooth lens model and an extended source. \rev{Fig. \ref{frvswl} shows the variation in flux ratio in SDSS J0924+0219 with wavelength.}

\begin{figure}  
\caption{Microlensing magnification map associated with image D. We zoom over a region of 50$\times$50 $\eta_0$. The colour scale corresponds to the the micro(de)magnification of a gaussian source of half-light radius $R_{1/2} = $ 0.06 $\eta_0$ (=1.8$\times10^{15}\,$cm = 5.8$\times10^{-4}\,$pc). By convention,  micro-magnification $\Delta m > 0$ correspond to the bright regions, and demagnification correspond to $\Delta m < 0$ as indicated by the colorbar. Several regions with demagnification stronger than $\lvert \Delta m \rvert = 2.7$\,mag (i.e. corresponding to a flux ratio 1:12 as observed in image $D$ over the last decade) are detected. Those regions sometimes extend over several microlensing Einstein radii $\eta_0$.} 

\hspace{-0.01cm}\includegraphics[width=9cm]{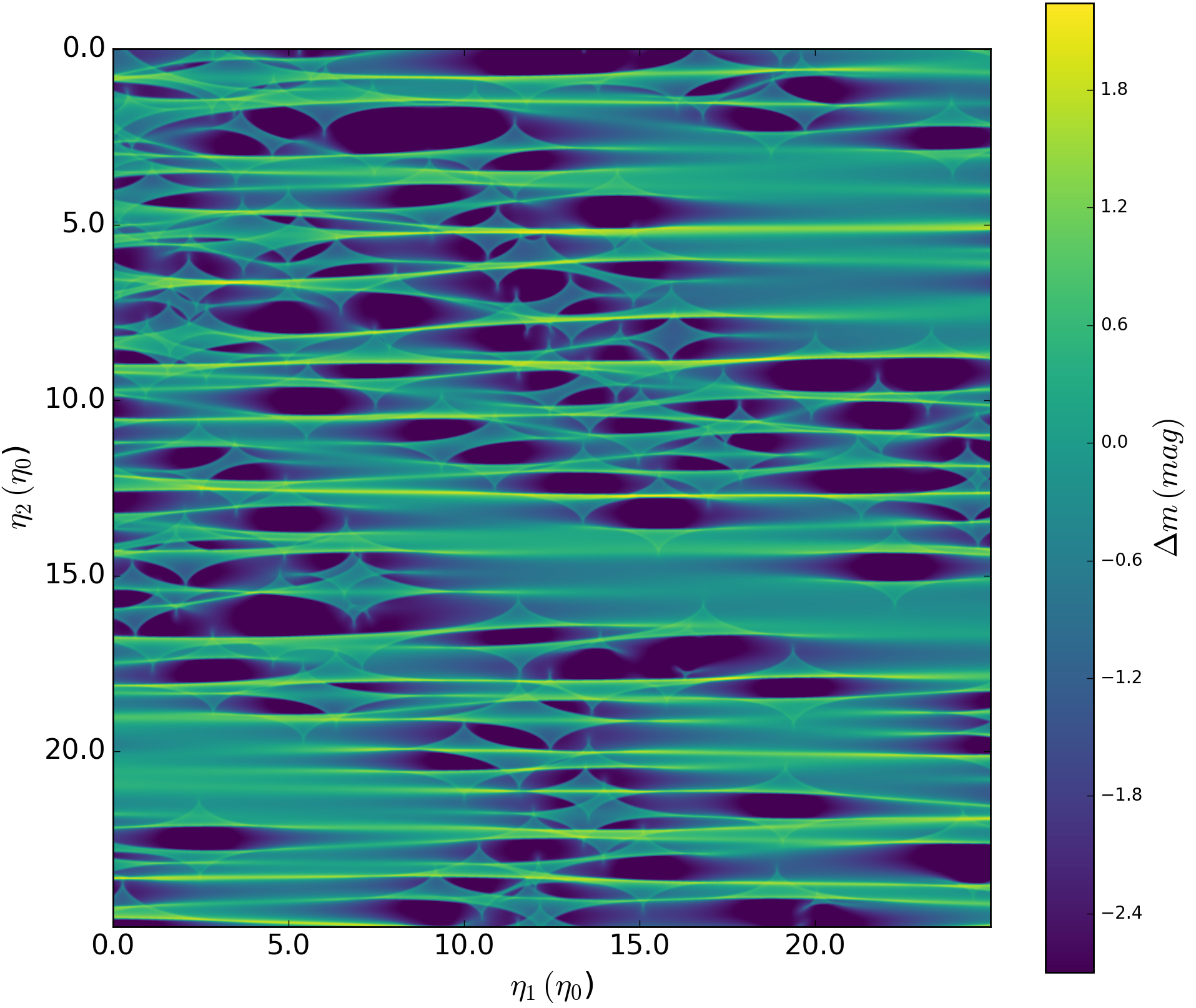}

\label{fig:ML}
\end{figure}

These results are consistent with the theory that the optical flux anomaly can be attributed to microlensing, a conclusion which is also consistent with previous studies of this object \citep{Keeton2006, Morgan2006}. As mentioned earlier, 15 years seems to be a long time for microlensing to persist. {\rev {To test that such a long period of demagnification is plausible, we simulated a microlensing magnification map corresponding to image D using the ray-tracing tree code developed by \cite{Wambsganss2001}. We built a map with 100 microlensing Einstein radius $\eta_0$ side-length and 0.005 $\eta_0$ / pixel resolution. For SDSS J0924+0219, \cite{Mosquera2011TheQuasars} calculate $\eta_0 \sim 3.2\times10^{16}$\, cm for an average microlens mass $<M/M_\odot>=0.3$, and a transverse velocity of $v_{\rm{trans}}=$0.049\,$\eta_0$/year. We use convergence and shear values at the position of image $D$ as derived by \cite{Keeton2006} (i.e. $(\kappa, \gamma)= (0.48, 0.57)$), a fraction of surface density in form of stars of $\alpha = 0.07$. This value of dark matter fraction is at the low side of values derived from microlensing at a projected distance of $\sim$ 1 Einstein radius from the lens center \citep{Jimenez2015}. To account for the optical-source size, we convolve the map by a gaussian profile with half-light radius $R_{1/2} = 0.02 \eta_0$, correponding to the $I-$band magnitude accretion disc source size expected in the framework of the standard accretion disc model \citep{Mosquera2011TheQuasars}. Figure~\ref{fig:ML} shows a fraction of this map where multiple regions demagnified by at least a factor 12 are observed. Those regions extend over several $\eta_0$, which translates to periods of demagnification that can last over 20 to 60 years. This result remains qualitatively unchanged if we increase the source size by a factor of 3, to account for the fact the the optical continuum emission may be larger than expected by the standard model of accretion \citep{Morgan2010TheRelation,Edelson2015SPACE5548,Hutsemekers2020SpatiallyJ081830.46+060138.0}.}}

The overall importance of observing systems like SDSS~J0924+0219 is that such observations contribute to the slow increase in the number of four-image strong lenses with available fluxes at optical wavelengths as well as wavelengths where the sources are large enough to be immune to microlensing. \citet{Xu2015HowAnomalies} considered the then best available sample of radio lenses and studied the effect of CDM substructure on each lens of the sample by adding substructure using Aquarius and Phoenix simulations to the macro models of the lens and examining the resulting flux ratio distributions, but like their predecessors was limited to the then available sample of $\sim$10 systems. \rev{The joint analysis of radio studies of radio quiet quasars and submillimetre/extended emission line studies is likely to be useful in investigating the physics of the radio emission and its relation to the molecular gas component (and possibly also substructure, with availability of the higher resolution submillimetre data).}

\newpage

\section*{Acknowledgements}
 The Karl G. Jansky Very Large Array is operated by the US National Radio Astronomy Observatory. NRAO is a facility of the National Science Foundation operated under cooperative agreement by Associated Universities, Inc. This paper makes use of the following ALMA data: ADS/JAO.ALMA2018.1.01447.S. ALMA is a partnership of ESO (representing its member states), NSF (USA) and NINS (Japan), together with NRC (Canada), MOST and ASIAA (Taiwan), and KASI (Republic of Korea), in cooperation with the Republic of Chile. The Joint ALMA Observatory is operated by ESO, AUI/NRAO and NAOJ. The authors would like to thank Ian Browne and an anonymous referee for their comments on the paper, Justin Spilker for his help with {\sc visilens} and George Bendo for his assistance with the ALMA data analysis.





\bibliographystyle{mnras}
\bibliography{0924bib_n, references} 

\begin{thebibliography}{}
\makeatletter
\relax
\def\mn@urlcharsother{\let\do\@makeother \do\$\do\&\do\#\do\^\do\_\do\%\do\~}
\def\mn@doi{\begingroup\mn@urlcharsother \@ifnextchar [ {\mn@doi@}
  {\mn@doi@[]}}
\def\mn@doi@[#1]#2{\def\@tempa{#1}\ifx\@tempa\@empty \href
  {http://dx.doi.org/#2} {doi:#2}\else \href {http://dx.doi.org/#2} {#1}\fi
  \endgroup}
\def\mn@eprint#1#2{\mn@eprint@#1:#2::\@nil}
\def\mn@eprint@arXiv#1{\href {http://arxiv.org/abs/#1} {{\tt arXiv:#1}}}
\def\mn@eprint@dblp#1{\href {http://dblp.uni-trier.de/rec/bibtex/#1.xml}
  {dblp:#1}}
\def\mn@eprint@#1:#2:#3:#4\@nil{\def\@tempa {#1}\def\@tempb {#2}\def\@tempc
  {#3}\ifx \@tempc \@empty \let \@tempc \@tempb \let \@tempb \@tempa \fi \ifx
  \@tempb \@empty \def\@tempb {arXiv}\fi \@ifundefined
  {mn@eprint@\@tempb}{\@tempb:\@tempc}{\expandafter \expandafter \csname
  mn@eprint@\@tempb\endcsname \expandafter{\@tempc}}}

\bibitem[\protect\citeauthoryear{Alloin, Guilloteau, Barvainis, Antonucci  \&
  Tacconi}{Alloin et~al.}{1997}]{Alloin1997TheEmission}
Alloin D.,  Guilloteau S.,  Barvainis R.,  Antonucci R.,   Tacconi L.,  1997,
  Astronomy and Astrophysics, 321, 24

\bibitem[\protect\citeauthoryear{Amara, Metcalf, Cox  \& Ostriker}{Amara
  et~al.}{2006}]{Amara2006SimulationsSubstructure}
Amara A.,  Metcalf R.~B.,  Cox T.~J.,   Ostriker J.~P.,  2006, \mn@doi [Monthly
  Notices of the Royal Astronomical Society]
  {10.1111/j.1365-2966.2006.10053.x}, 367, 1367

\bibitem[\protect\citeauthoryear{Anh, Boone, Hoai, Nhung, Wei{\ss}, Kneib,
  Beelen  \& Salom{\'{e}}}{Anh et~al.}{2013}]{Anh2013ResolvingRXJ0911.4+0551}
Anh P.~T.,  Boone F.,  Hoai D.~T.,  Nhung P.~T.,  Wei{\ss} A.,  Kneib J.-P.,
  Beelen A.,   Salom{\'{e}} P.,  2013, \mn@doi [Astronomy {\&} Astrophysics]
  {10.1051/0004-6361/201321363}, 552, L12

\bibitem[\protect\citeauthoryear{{Aravena} et~al.,}{{Aravena}
  et~al.}{2016}]{ara16}
{Aravena} M.,  et~al., 2016, \mn@doi [Monthly Notices of the Royal Astronomical
  Society] {10.1093/mnras/stw275}, \href
  {https://ui.adsabs.harvard.edu/abs/2016MNRAS.457.4406A} {457, 4406}

\bibitem[\protect\citeauthoryear{Atek et~al.,}{Atek
  et~al.}{2015}]{Atek2015NewA2744}
Atek H.,  et~al., 2015, \mn@doi [The Astrophysical Journal]
  {10.1088/0004-637X/800/1/18}, 800, 18

\bibitem[\protect\citeauthoryear{Beelen et~al.,}{Beelen
  et~al.}{2004}]{Beelen2004StarburstJ1409+5628}
Beelen A.,  et~al., 2004, \mn@doi [Astronomy {\&} Astrophysics]
  {10.1051/0004-6361:20040318}, 423, 441

\bibitem[\protect\citeauthoryear{{Bonzini} et~al.,}{{Bonzini}
  et~al.}{2015}]{bon15}
{Bonzini} M.,  et~al., 2015, \mn@doi [Monthly Notices of the Royal Astronomical
  Society] {10.1093/mnras/stv1675}, \href
  {https://ui.adsabs.harvard.edu/abs/2015MNRAS.453.1079B} {453, 1079}

\bibitem[\protect\citeauthoryear{Brusa et~al.,}{Brusa
  et~al.}{2018}]{Brusa2018MolecularALMA}
Brusa M.,  et~al., 2018, \mn@doi [Astronomy {\&} Astrophysics]
  {10.1051/0004-6361/201731641}, 612, A29

\bibitem[\protect\citeauthoryear{Burud et~al.,}{Burud
  et~al.}{2002}]{Burud2002An2149-2745}
Burud I.,  et~al., 2002, \mn@doi [Astronomy {\&} Astrophysics]
  {10.1051/0004-6361:20011731}, 383, 71

\bibitem[\protect\citeauthoryear{Chang \& Refsdal}{Chang \&
  Refsdal}{1979}]{Chang1979FluxPath}
Chang K.,  Refsdal S.,  1979, \mn@doi [Nature] {10.1038/282561a0}, 282, 561

\bibitem[\protect\citeauthoryear{Chiba}{Chiba}{2002}]{Chiba2002ProbingGalaxies}
Chiba M.,  2002, \mn@doi [The Astrophysical Journal] {10.1086/324493}, 565, 17

\bibitem[\protect\citeauthoryear{Chiba, Minezaki, Kashikawa, Kataza  \&
  Inoue}{Chiba et~al.}{2005}]{Chiba2005SubaruLensing}
Chiba M.,  Minezaki T.,  Kashikawa N.,  Kataza H.,   Inoue K.~T.,  2005,
  \mn@doi [The Astrophysical Journal] {10.1086/430403}, 627, 53

\bibitem[\protect\citeauthoryear{{Condon}, {Kellermann}, {Kimball},
  {Ivezi{\'c}}  \& {Perley}}{{Condon} et~al.}{2013}]{con13}
{Condon} J.~J.,  {Kellermann} K.~I.,  {Kimball} A.~E.,  {Ivezi{\'c}} {\v{Z}}.,
   {Perley} R.~A.,  2013, \mn@doi [The Astrophysical Journal]
  {10.1088/0004-637X/768/1/37}, \href
  {https://ui.adsabs.harvard.edu/abs/2013ApJ...768...37C} {768, 37}

\bibitem[\protect\citeauthoryear{{Croton} et~al.,}{{Croton}
  et~al.}{2006}]{cro06}
{Croton} D.~J.,  et~al., 2006, \mn@doi [Monthly Notices of the Royal
  Astronomical Society] {10.1111/j.1365-2966.2005.09675.x}, \href
  {https://ui.adsabs.harvard.edu/abs/2006MNRAS.365...11C} {365, 11}

\bibitem[\protect\citeauthoryear{Dalal \& Kochanek}{Dalal \&
  Kochanek}{2002}]{Dalal2002}
Dalal N.,  Kochanek C.~S.,  2002, \mn@doi [The Astrophysical Journal]
  {10.1086/340303}, 572, 25

\bibitem[\protect\citeauthoryear{{Deane}, {Heywood}, {Rawlings}  \&
  {Marshall}}{{Deane} et~al.}{2013}]{dea13}
{Deane} R.~P.,  {Heywood} I.,  {Rawlings} S.,   {Marshall} P.~J.,  2013,
  \mn@doi [Monthly Notices of the Royal Astronomical Society]
  {10.1093/mnras/stt957}, \href
  {https://ui.adsabs.harvard.edu/abs/2013MNRAS.434...23D} {434, 23}

\bibitem[\protect\citeauthoryear{Downes, Neri, Wiklind, Wilner  \&
  Shaver}{Downes et~al.}{1999}]{Downes1999Detection3.9}
Downes D.,  Neri R.,  Wiklind T.,  Wilner D.~J.,   Shaver P.~A.,  1999, \mn@doi
  [The Astrophysical Journal] {10.1086/311896}, 513, L1

\bibitem[\protect\citeauthoryear{Edelson et~al.,}{Edelson
  et~al.}{2015}]{Edelson2015SPACE5548}
Edelson R.,  et~al., 2015, \mn@doi [The Astrophysical Journal]
  {10.1088/0004-637X/806/1/129}, 806, 129

\bibitem[\protect\citeauthoryear{Eigenbrod, Courbin, Dye, Meylan, Sluse,
  Vuissoz  \& Magain}{Eigenbrod et~al.}{2006}]{Eigenbrod2006COSMOGRAIL:Rings}
Eigenbrod A.,  Courbin F.,  Dye S.,  Meylan G.,  Sluse D.,  Vuissoz C.,
  Magain P.,  2006, \mn@doi [Astronomy {\&} Astrophysics]
  {10.1051/0004-6361:20054423}, 451, 747

\bibitem[\protect\citeauthoryear{Fadely \& Keeton}{Fadely \&
  Keeton}{2011}]{Fadely2011Near-infraredQuasars}
Fadely R.,  Keeton C.~R.,  2011, \mn@doi [The Astronomical Journal]
  {10.1088/0004-6256/141/3/101}, 141, 101

\bibitem[\protect\citeauthoryear{Fadely \& Keeton}{Fadely \&
  Keeton}{2012}]{Fadely2012Substructure0435-1223}
Fadely R.,  Keeton C.~R.,  2012, \mn@doi [Monthly Notices of the Royal
  Astronomical Society] {10.1111/j.1365-2966.2011.19729.x}, 419, 936

\bibitem[\protect\citeauthoryear{Fassnacht et~al.,}{Fassnacht
  et~al.}{1998}]{fass}
Fassnacht C.~D.,  et~al., 1998, \mn@doi [The Astronomical Journal]
  {10.1086/300724}, 117, 658

\bibitem[\protect\citeauthoryear{Faure, Sluse, Cantale, Tewes, Courbin, Durrer
  \& Meylan}{Faure et~al.}{2011}]{Faure2011}
Faure C.,  Sluse D.,  Cantale N.,  Tewes M.,  Courbin F.,  Durrer P.,   Meylan
  G.,  2011, \mn@doi [Astronomy {\&} Astrophysics]
  {10.1051/0004-6361/201117787}, 536, A29

\bibitem[\protect\citeauthoryear{Feruglio et~al.,}{Feruglio
  et~al.}{2018}]{Feruglio2018TheALMA}
Feruglio C.,  et~al., 2018, \mn@doi [Astronomy {\&} Astrophysics]
  {10.1051/0004-6361/201833174}, 619, A39

\bibitem[\protect\citeauthoryear{Floyd, Bate  \& Webster}{Floyd
  et~al.}{2009}]{Floyd2009TheJ0924+0219}
Floyd D.~J.,  Bate N.~F.,   Webster R.~L.,  2009, \mn@doi [Monthly Notices of
  the Royal Astronomical Society] {10.1111/j.1365-2966.2009.15045.x}, 398, 233

\bibitem[\protect\citeauthoryear{{Gaia Collaboration}}{{Gaia
  Collaboration}}{2018}]{gaia}
{Gaia Collaboration} e.~a.,  2018, astro-ph/1804.09365

\bibitem[\protect\citeauthoryear{{Geach}, {Ivison}, {Dye}  \& {Oteo}}{{Geach}
  et~al.}{2018}]{gea18}
{Geach} J.~E.,  {Ivison} R.~J.,  {Dye} S.,   {Oteo} I.,  2018, \mn@doi [\apjl]
  {10.3847/2041-8213/aae375}, \href
  {https://ui.adsabs.harvard.edu/abs/2018ApJ...866L..12G} {866, L12}

\bibitem[\protect\citeauthoryear{Gilman, Birrer, Treu, Nierenberg  \&
  Benson}{Gilman et~al.}{2019}]{Gilman2019ProbingHaloes}
Gilman D.,  Birrer S.,  Treu T.,  Nierenberg A.,   Benson A.,  2019, \mn@doi
  [Monthly Notices of the Royal Astronomical Society] {10.1093/mnras/stz1593},
  487, 5721

\bibitem[\protect\citeauthoryear{{Hartley}, {Jackson}, {Sluse}, {Stacey}  \&
  {Vives-Arias}}{{Hartley} et~al.}{2019}]{har19}
{Hartley} P.,  {Jackson} N.,  {Sluse} D.,  {Stacey} H.~R.,   {Vives-Arias} H.,
  2019, \mn@doi [Monthly Notices of the Royal Astronomical Society]
  {10.1093/mnras/stz510}, \href
  {https://ui.adsabs.harvard.edu/abs/2019MNRAS.485.3009H} {485, 3009}

\bibitem[\protect\citeauthoryear{{Herrera Ruiz}, {Middelberg}, {Norris}  \&
  {Maini}}{{Herrera Ruiz} et~al.}{2016}]{her16}
{Herrera Ruiz} N.,  {Middelberg} E.,  {Norris} R.~P.,   {Maini} A.,  2016,
  \mn@doi [Astronomy & Astrophysics] {10.1051/0004-6361/201628302}, \href
  {https://ui.adsabs.harvard.edu/abs/2016A&A...589L...2H} {589, L2}

\bibitem[\protect\citeauthoryear{Hezaveh, Dalal, Holder, Kuhlen, Marrone,
  Murray  \& Vieira}{Hezaveh et~al.}{2013a}]{Hezaveh2013DarkGalaxies}
Hezaveh Y.,  Dalal N.,  Holder G.,  Kuhlen M.,  Marrone D.,  Murray N.,
  Vieira J.,  2013a, \mn@doi [The Astrophysical Journal]
  {10.1088/0004-637X/767/1/9}, 767, 9

\bibitem[\protect\citeauthoryear{Hezaveh et~al.,}{Hezaveh
  et~al.}{2013b}]{Hezaveh2013AlmaGalaxies}
Hezaveh Y.~D.,  et~al., 2013b, \mn@doi [The Astrophysical Journal]
  {10.1088/0004-637X/767/2/132}, 767, 132

\bibitem[\protect\citeauthoryear{Hezaveh et~al.,}{Hezaveh
  et~al.}{2016}]{Hezaveh2016DetectionSdp.81}
Hezaveh Y.~D.,  et~al., 2016, \mn@doi [The Astrophysical Journal]
  {10.3847/0004-637x/823/1/37}, 823, 37

\bibitem[\protect\citeauthoryear{Hodge et~al.,}{Hodge
  et~al.}{2016}]{Hodge2016Kiloparsec-ScaleGalaxies}
Hodge J.~A.,  et~al., 2016, \mn@doi [The Astrophysical Journal]
  {10.3847/1538-4357/833/1/103}, 833, 103

\bibitem[\protect\citeauthoryear{Hodge et~al.,}{Hodge
  et~al.}{2019}]{Hodge2019ALMAGalaxies}
Hodge J.~A.,  et~al., 2019, \mn@doi [The Astrophysical Journal]
  {10.3847/1538-4357/ab1846}, 876, 130

\bibitem[\protect\citeauthoryear{H{\"{o}}gbom}{H{\"{o}}gbom}{1974}]{Hogbom1974ApertureBaselines}
H{\"{o}}gbom J.~A.,  1974, Astronomy and Astrophysics Supplement Series, 15,
  417

\bibitem[\protect\citeauthoryear{Hsueh, Fassnacht, Vegetti, McKean, Spingola,
  Auger, Koopmans  \& Lagattuta}{Hsueh et~al.}{2016}]{hsueh}
Hsueh J.~W.,  Fassnacht C.~D.,  Vegetti S.,  McKean J.~P.,  Spingola C.,  Auger
  M.~W.,  Koopmans L.~V.,   Lagattuta D.~J.,  2016, \mn@doi [Monthly Notices of
  the Royal Astronomical Society: Letters] {10.1093/mnrasl/slw146}, 463, L51

\bibitem[\protect\citeauthoryear{Hutsem{\'{e}}kers, Sluse  \&
  Kumar}{Hutsem{\'{e}}kers
  et~al.}{2020}]{Hutsemekers2020SpatiallyJ081830.46+060138.0}
Hutsem{\'{e}}kers D.,  Sluse D.,   Kumar P.,  2020, \mn@doi [Astronomy {\&}
  Astrophysics] {10.1051/0004-6361/201936973}, 633, A101

\bibitem[\protect\citeauthoryear{Inada et~al.,}{Inada et~al.}{2003}]{Inada2003}
Inada N.,  et~al., 2003, \mn@doi [The Astronomical Journal] {10.1086/375906},
  126, 666

\bibitem[\protect\citeauthoryear{Irwin, Webster, Hewett, Corrigan  \&
  Jedrzejewski}{Irwin et~al.}{1989}]{Irwin1989PhotometricEvent}
Irwin M.~J.,  Webster R.~L.,  Hewett P.~C.,  Corrigan R.~T.,   Jedrzejewski
  R.~I.,  1989, \mn@doi [The Astronomical Journal] {10.1086/115272}, 98

\bibitem[\protect\citeauthoryear{Ivison et~al.,}{Ivison
  et~al.}{2010}]{Ivison2010TheHerschel}
Ivison R.~J.,  et~al., 2010, \mn@doi [Astronomy {\&} Astrophysics]
  {10.1051/0004-6361/201014552}, 518, 1

\bibitem[\protect\citeauthoryear{{Jackson}}{{Jackson}}{2011}]{jac11}
{Jackson} N.,  2011, \mn@doi [The Astrophysical Journal Letters]
  {10.1088/2041-8205/739/1/L28}, \href
  {https://ui.adsabs.harvard.edu/abs/2011ApJ...739L..28J} {739, L28}

\bibitem[\protect\citeauthoryear{{Jackson}, {Tagore}, {Roberts}, {Sluse},
  {Stacey}, {Vives-Arias}, {Wucknitz}  \& {Volino}}{{Jackson}
  et~al.}{2015}]{jac15}
{Jackson} N.,  {Tagore} A.~S.,  {Roberts} C.,  {Sluse} D.,  {Stacey} H.,
  {Vives-Arias} H.,  {Wucknitz} O.,   {Volino} F.,  2015, \mn@doi [Monthly
  Notices of the Royal Astronomical Society] {10.1093/mnras/stv1982}, \href
  {https://ui.adsabs.harvard.edu/abs/2015MNRAS.454..287J} {454, 287}

\bibitem[\protect\citeauthoryear{Jarvis et~al.,}{Jarvis
  et~al.}{2019}]{Jarvis2019PrevalenceQuasars}
Jarvis M.~E.,  et~al., 2019, \mn@doi [Monthly Notices of the Royal Astronomical
  Society] {10.1093/mnras/stz556}, 485, 2710

\bibitem[\protect\citeauthoryear{{Jim{\'e}nez-Vicente}, {Mediavilla},
  {Kochanek}  \& {Mu{\~n}oz}}{{Jim{\'e}nez-Vicente} et~al.}{2015}]{Jimenez2015}
{Jim{\'e}nez-Vicente} J.,  {Mediavilla} E.,  {Kochanek} C.~S.,   {Mu{\~n}oz}
  J.~A.,  2015, \mn@doi [The Astrophysical Journal]
  {10.1088/0004-637X/799/2/149}, \href
  {https://ui.adsabs.harvard.edu/abs/2015ApJ...799..149J} {799, 149}

\bibitem[\protect\citeauthoryear{Keeton, Burles, Schechter  \&
  Wambsganss}{Keeton et~al.}{2006}]{Keeton2006}
Keeton C.~R.,  Burles S.,  Schechter P.~L.,   Wambsganss J.,  2006, \mn@doi
  [The Astrophysical Journal] {10.1086/499264}, 639, 1

\bibitem[\protect\citeauthoryear{Kratzer, Richards, Goldberg, Oguri, Kochanek,
  Hodge, Becker  \& Inada}{Kratzer
  et~al.}{2011}]{Kratzer2011ANALYZINGJ1029+2623}
Kratzer R.~M.,  Richards G.~T.,  Goldberg D.~M.,  Oguri M.,  Kochanek C.~S.,
  Hodge J.~A.,  Becker R.~H.,   Inada N.,  2011, \mn@doi [The Astrophysical
  Journal] {10.1088/2041-8205/728/1/L18}, 728, L18

\bibitem[\protect\citeauthoryear{MacLeod et~al.,}{MacLeod
  et~al.}{2015}]{MacLeod2015}
MacLeod C.~L.,  et~al., 2015, \mn@doi [The Astrophysical Journal]
  {10.1088/0004-637X/806/2/258}, 806, 258

\bibitem[\protect\citeauthoryear{{Maiolino} et~al.,}{{Maiolino}
  et~al.}{2017}]{mai17}
{Maiolino} R.,  et~al., 2017, \mn@doi [\nat] {10.1038/nature21677}, \href
  {https://ui.adsabs.harvard.edu/abs/2017Natur.544..202M} {544, 202}

\bibitem[\protect\citeauthoryear{Mao \& Schneider}{Mao \&
  Schneider}{1998}]{Mao1998a}
Mao S.,  Schneider P.,  1998, \mn@doi [Monthly Notices of the Royal
  Astronomical Society] {10.1046/j.1365-8711.1998.01319.x}, 295, 587

\bibitem[\protect\citeauthoryear{Mart{\'{i}}-Vidal, Vlemmings, Muller  \&
  Casey}{Mart{\'{i}}-Vidal et~al.}{2014}]{Marti-Vidal2014UVMULTIFIT:Data}
Mart{\'{i}}-Vidal I.,  Vlemmings W.~H.,  Muller S.,   Casey S.,  2014, \mn@doi
  [Astronomy {\&} Astrophysics] {10.1051/0004-6361/201322633}, 563, A136

\bibitem[\protect\citeauthoryear{Mcmullin, Waters, Schiebel, Young  \&
  Golap}{Mcmullin et~al.}{2007}]{Mcmullin2007CASAApplications}
Mcmullin J.~P.,  Waters B.,  Schiebel D.,  Young W.,   Golap K.,  2007, in
  Astronomical Data Analysis Software and Systems XVI ASP Conference Series.
  p.~127

\bibitem[\protect\citeauthoryear{Metcalf \& Madau}{Metcalf \&
  Madau}{2001}]{Metcalf2001CompoundHalos}
Metcalf R.~B.,  Madau P.,  2001, \mn@doi [The Astrophysical Journal]
  {10.1086/323695}, 563, 9

\bibitem[\protect\citeauthoryear{Metcalf \& Zhao}{Metcalf \&
  Zhao}{2002}]{Metcalf2002FluxLenses}
Metcalf R.~B.,  Zhao H.,  2002, \mn@doi [The Astrophysical Journal]
  {10.1086/339798}, 567, L5

\bibitem[\protect\citeauthoryear{Miranda \& Macci{\`{o}}}{Miranda \&
  Macci{\`{o}}}{2007}]{Miranda2007ConstrainingLensing}
Miranda M.,  Macci{\`{o}} A.~V.,  2007, \mn@doi [Monthly Notices of the Royal
  Astronomical Society] {10.1111/j.1365-2966.2007.12440.x}, 382, 1225

\bibitem[\protect\citeauthoryear{M{\"{o}}ller, Hewett  \& Blain}{M{\"{o}}ller
  et~al.}{2003}]{Moller2003DiscsH0}
M{\"{o}}ller O.,  Hewett P.,   Blain A.~W.,  2003, \mn@doi [Monthly Notices of
  the Royal Astronomical Society] {10.1046/j.1365-8711.2003.06758.x}, 345, 1

\bibitem[\protect\citeauthoryear{Morgan, Kochanek, Morgan  \& Falco}{Morgan
  et~al.}{2006}]{Morgan2006}
Morgan C.~W.,  Kochanek C.~S.,  Morgan N.~D.,   Falco E.~E.,  2006, \mn@doi
  [The Astrophysical Journal] {10.1086/505569}, 647, 874

\bibitem[\protect\citeauthoryear{Morgan, Kochanek, Morgan  \& Falco}{Morgan
  et~al.}{2010}]{Morgan2010TheRelation}
Morgan C.~W.,  Kochanek C.~S.,  Morgan N.~D.,   Falco E.~E.,  2010, \mn@doi
  [The Astrophysical Journal] {10.1088/0004-637X/712/2/1129}, 712, 1129

\bibitem[\protect\citeauthoryear{Mosquera \& Kochanek}{Mosquera \&
  Kochanek}{2011}]{Mosquera2011TheQuasars}
Mosquera A.~M.,  Kochanek C.~S.,  2011, \mn@doi [The Astrophysical Journal]
  {10.1088/0004-637X/738/1/96}, 738, 96

\bibitem[\protect\citeauthoryear{Moustakas \& Metcalf}{Moustakas \&
  Metcalf}{2003}]{Moustakas2003DetectingLenses}
Moustakas L.~A.,  Metcalf R.~B.,  2003, \mn@doi [Monthly Notices of the Royal
  Astronomical Society] {10.1046/j.1365-8711.2003.06055.x}, 339, 607

\bibitem[\protect\citeauthoryear{Nierenberg et~al.,}{Nierenberg
  et~al.}{2017}]{Nierenberg2017ProbingGrism}
Nierenberg A.~M.,  et~al., 2017, \mn@doi [Monthly Notices of the Royal
  Astronomical Society] {10.1093/mnras/stx1400}, 471, 2224

\bibitem[\protect\citeauthoryear{Nierenberg et~al.,}{Nierenberg
  et~al.}{2020}]{Nierenberg2020DoubleGrism}
Nierenberg A.~M.,  et~al., 2020, \mn@doi [Monthly Notices of the Royal
  Astronomical Society] {https://doi.org/10.1093/mnras/stz3588}, 492, 5314

\bibitem[\protect\citeauthoryear{{Padovani} et~al.,}{{Padovani}
  et~al.}{2017}]{pad17}
{Padovani} P.,  et~al., 2017, \mn@doi [The Astronomy and Astrophysics Review]
  {10.1007/s00159-017-0102-9}, \href
  {https://ui.adsabs.harvard.edu/abs/2017A&ARv..25....2P} {25, 2}

\bibitem[\protect\citeauthoryear{Paraficz et~al.,}{Paraficz
  et~al.}{2018}]{Paraficz2018ALMAGalaxy}
Paraficz D.,  et~al., 2018, \mn@doi [Astronomy {\&} Astrophysics]
  {10.1051/0004-6361/201731250}, 613, A34

\bibitem[\protect\citeauthoryear{Peeples, Schechter  \& Wambsganss}{Peeples
  et~al.}{2004}]{2004AAS...205.2806P}
Peeples M.~S.,  Schechter P.~L.,   Wambsganss J.~K.,  2004, in BAAS. p.~1392

\bibitem[\protect\citeauthoryear{Poindexter, Morgan  \& Kochanek}{Poindexter
  et~al.}{2008}]{Poindexter2008TheDisk}
Poindexter S.,  Morgan N.,   Kochanek C.~S.,  2008, \mn@doi [The Astrophysical
  Journal] {10.1086/524190}, 673, 34

\bibitem[\protect\citeauthoryear{Polletta, Nesvadba, Neri, Omont, Berta  \&
  Bergeron}{Polletta et~al.}{2011}]{Polletta2011DiskZ3.4}
Polletta M.,  Nesvadba N.~P.,  Neri R.,  Omont A.,  Berta S.,   Bergeron J.,
  2011, \mn@doi [Astronomy {\&} Astrophysics] {10.1051/0004-6361/201116446},
  533, A20

\bibitem[\protect\citeauthoryear{Riechers, Walter, Carilli  \& Lewis}{Riechers
  et~al.}{2009a}]{Riechers2009Imaging08279+5255}
Riechers D.~A.,  Walter F.,  Carilli C.~L.,   Lewis G.~F.,  2009a, \mn@doi [The
  Astrophysical Journal] {10.1088/0004-637X/690/1/463}, 690, 463

\bibitem[\protect\citeauthoryear{Riechers et~al.,}{Riechers
  et~al.}{2009b}]{Riechers2009ImagingReionization}
Riechers D.~A.,  et~al., 2009b, \mn@doi [The Astrophysical Journal]
  {10.1088/0004-637X/703/2/1338}, 703, 1338

\bibitem[\protect\citeauthoryear{{Riechers} et~al.,}{{Riechers}
  et~al.}{2011}]{rie11}
{Riechers} D.~A.,  et~al., 2011, \mn@doi [The Astrophysical Journal Letters]
  {10.1088/2041-8205/739/1/L32}, \href
  {https://ui.adsabs.harvard.edu/abs/2011ApJ...739L..32R} {739, L32}

\bibitem[\protect\citeauthoryear{{Rybak}, {Vegetti}, {McKean}, {Andreani}  \&
  {White}}{{Rybak} et~al.}{2015}]{ryb15}
{Rybak} M.,  {Vegetti} S.,  {McKean} J.~P.,  {Andreani} P.,   {White} S.~D.~M.,
   2015, \mn@doi [Monthly Notices of the Royal Astronomical Society]
  {10.1093/mnrasl/slv092}, \href
  {https://ui.adsabs.harvard.edu/abs/2015MNRAS.453L..26R} {453, L26}

\bibitem[\protect\citeauthoryear{Saha, Williams  \& Ferreras}{Saha
  et~al.}{2007}]{Saha2007MesoStructureSystems}
Saha P.,  Williams L. L.~R.,   Ferreras I.,  2007, \mn@doi [The Astrophysical
  Journal] {10.1086/518083}, 663, 29

\bibitem[\protect\citeauthoryear{Smit et~al.,}{Smit
  et~al.}{2018}]{Smit2018Rotation6.8}
Smit R.,  et~al., 2018, \mn@doi [Nature] {10.1038/nature24631}, 553, 178

\bibitem[\protect\citeauthoryear{{Solomon} \& {Vanden Bout}}{{Solomon} \&
  {Vanden Bout}}{2005}]{sol05}
{Solomon} P.~M.,  {Vanden Bout} P.~A.,  2005, \mn@doi [Annual Review of
  Astronomy and Astrophysics] {10.1146/annurev.astro.43.051804.102221}, \href
  {https://ui.adsabs.harvard.edu/abs/2005ARA&A..43..677S} {43, 677}

\bibitem[\protect\citeauthoryear{Spilker et~al.,}{Spilker
  et~al.}{2016}]{Spilker2016ALMAREDSHIFTS}
Spilker J.~S.,  et~al., 2016, \mn@doi [The Astrophysical Journal]
  {10.3847/0004-637x/826/2/112}, 826, 112

\bibitem[\protect\citeauthoryear{{Spingola} et~al.,}{{Spingola}
  et~al.}{2019}]{spi19}
{Spingola} C.,  et~al., 2019, arXiv e-prints, \href
  {https://ui.adsabs.harvard.edu/abs/2019arXiv190506363S} {p. arXiv:1905.06363}

\bibitem[\protect\citeauthoryear{{Stacey} et~al.,}{{Stacey}
  et~al.}{2018}]{sta18}
{Stacey} H.~R.,  et~al., 2018, \mn@doi [Monthly Notices of the Royal
  Astronomical Society] {10.1093/mnras/sty458}, \href
  {https://ui.adsabs.harvard.edu/abs/2018MNRAS.476.5075S} {476, 5075}

\bibitem[\protect\citeauthoryear{Stacey et~al.,}{Stacey
  et~al.}{2019}]{Stacey2019LoTSS/HETDEX:Quasars}
Stacey H.~R.,  et~al., 2019, \mn@doi [Astronomy {\&} Astrophysics]
  {10.1051/0004-6361/201833967}, 622, A18

\bibitem[\protect\citeauthoryear{Sugai, Kawai, Shimono, Hattori, Kosugi,
  Kashikawa, Inoue  \& Chiba}{Sugai
  et~al.}{2007}]{Sugai2007IntegralSubstructures}
Sugai H.,  Kawai A.,  Shimono A.,  Hattori T.,  Kosugi G.,  Kashikawa N.,
  Inoue K.~T.,   Chiba M.,  2007, \mn@doi [The Astrophysical Journal]
  {10.1086/513731}, 660, 1016

\bibitem[\protect\citeauthoryear{Treu}{Treu}{2010}]{Treu2010}
Treu T.,  2010, \mn@doi [Annual Review of Astronomy and Astrophysics]
  {10.1146/annurev-astro-081309-130924}, 48, 87

\bibitem[\protect\citeauthoryear{Tuan-Anh, Hoai, Nhung, Diep, Phuong, Thao  \&
  Darriulat}{Tuan-Anh et~al.}{2017}]{Tuan-Anh2017OnJ0911.4+0551}
Tuan-Anh P.,  Hoai D.~T.,  Nhung P.~T.,  Diep P.~N.,  Phuong N.~T.,  Thao
  N.~T.,   Darriulat P.,  2017, \mn@doi [Monthly Notices of the Royal
  Astronomical Society] {10.1093/mnras/stx212}, 467, 3513

\bibitem[\protect\citeauthoryear{Vegetti \& Koopmans}{Vegetti \&
  Koopmans}{2009}]{Vegetti2009BayesianGalaxies}
Vegetti S.,  Koopmans L.~V.,  2009, \mn@doi [Monthly Notices of the Royal
  Astronomical Society] {10.1111/j.1365-2966.2008.14005.x}, 392, 945

\bibitem[\protect\citeauthoryear{Vegetti, Czoske  \& Koopmans}{Vegetti
  et~al.}{2010a}]{Vegetti2010QuantifyingJ120602.09+514229.5}
Vegetti S.,  Czoske O.,   Koopmans L.~V.,  2010a, \mn@doi [Monthly Notices of
  the Royal Astronomical Society] {10.1111/j.1365-2966.2010.16952.x}, 407, 225

\bibitem[\protect\citeauthoryear{Vegetti, Koopmans, Bolton, Treu  \&
  Gavazzi}{Vegetti et~al.}{2010b}]{Vegetti2010DetectionImaging}
Vegetti S.,  Koopmans L.~V.,  Bolton A.,  Treu T.,   Gavazzi R.,  2010b,
  \mn@doi [Monthly Notices of the Royal Astronomical Society]
  {10.1111/j.1365-2966.2010.16865.x}, 408, 1969

\bibitem[\protect\citeauthoryear{{Venturini} \& {Solomon}}{{Venturini} \&
  {Solomon}}{2003}]{ven03}
{Venturini} S.,  {Solomon} P.~M.,  2003, \mn@doi [The Astrophysical Journal]
  {10.1086/375050}, \href
  {https://ui.adsabs.harvard.edu/abs/2003ApJ...590..740V} {590, 740}

\bibitem[\protect\citeauthoryear{Vuissoz et~al.,}{Vuissoz
  et~al.}{2007}]{Vuissoz2007COSMOGRAIL:J1650+4251}
Vuissoz C.,  et~al., 2007, \mn@doi [Astronomy {\&} Astrophysics]
  {10.1051/0004-6361:20065823}, 464, 845

\bibitem[\protect\citeauthoryear{Vuissoz et~al.,}{Vuissoz
  et~al.}{2008}]{Vuissoz2008COSMOGRAIL:J2033-4723}
Vuissoz C.,  et~al., 2008, \mn@doi [Astronomy {\&} Astrophysics]
  {10.1051/0004-6361:200809866}, 488, 481

\bibitem[\protect\citeauthoryear{Wambsganss}{Wambsganss}{2001}]{Wambsganss2001}
Wambsganss J.,  2001, \mn@doi [Publications of the Astronomical Society of
  Australia] {10.1071/AS01016}, 18, 207

\bibitem[\protect\citeauthoryear{{Warren} \& {Dye}}{{Warren} \&
  {Dye}}{2003}]{war03}
{Warren} S.~J.,  {Dye} S.,  2003, \mn@doi [The Astrophysical Journal]
  {10.1086/375132}, \href
  {https://ui.adsabs.harvard.edu/abs/2003ApJ...590..673W} {590, 673}

\bibitem[\protect\citeauthoryear{{White}, {Jarvis}, {Kalfountzou},
  {Hardcastle}, {Verma}, {Cao Orjales}  \& {Stevens}}{{White}
  et~al.}{2017}]{whi17}
{White} S.~V.,  {Jarvis} M.~J.,  {Kalfountzou} E.,  {Hardcastle} M.~J.,
  {Verma} A.,  {Cao Orjales} J.~M.,   {Stevens} J.,  2017, \mn@doi [Monthly
  Notices of the Royal Astronomical Society] {10.1093/mnras/stx284}, \href
  {https://ui.adsabs.harvard.edu/abs/2017MNRAS.468..217W} {468, 217}

\bibitem[\protect\citeauthoryear{Witt, Mao  \& Schechter}{Witt
  et~al.}{1995}]{Witt1995OnLenses}
Witt H.,  Mao S.,   Schechter P.,  1995, \mn@doi [The Astrophysical Journal]
  {10.1086/175499}, 443, 18

\bibitem[\protect\citeauthoryear{{Wucknitz} \& {Volino}}{{Wucknitz} \&
  {Volino}}{2008}]{wuc08}
{Wucknitz} O.,  {Volino} F.,  2008, arXiv e-prints, \href
  {https://ui.adsabs.harvard.edu/abs/2008arXiv0811.3421W} {p. arXiv:0811.3421}

\bibitem[\protect\citeauthoryear{Xu, Mao, Cooper, Gao, Frenk, Angulo  \&
  Helly}{Xu et~al.}{2012}]{xulos}
Xu D.~D.,  Mao S.,  Cooper A.~P.,  Gao L.,  Frenk C.~S.,  Angulo R.~E.,   Helly
  J.,  2012, \mn@doi [Monthly Notices of the Royal Astronomical Society]
  {10.1111/j.1365-2966.2012.20484.x}, 421, 2553

\bibitem[\protect\citeauthoryear{Xu, Sluse, Gao, Wang, Frenk, Mao, Schneider
  \& Springel}{Xu et~al.}{2015}]{Xu2015HowAnomalies}
Xu D.,  Sluse D.,  Gao L.,  Wang J.,  Frenk C.,  Mao S.,  Schneider P.,
  Springel V.,  2015, \mn@doi [Monthly Notices of the Royal Astronomical
  Society] {10.1093/mnras/stu2673}, 447, 3189

\bibitem[\protect\citeauthoryear{{Yang} et~al.,}{{Yang} et~al.}{2017}]{yan17}
{Yang} C.,  et~al., 2017, \mn@doi [Astronomy & Astrophysics]
  {10.1051/0004-6361/201731391}, \href
  {https://ui.adsabs.harvard.edu/abs/2017A&A...608A.144Y} {608, A144}

\bibitem[\protect\citeauthoryear{Yuan, Kewley, Swinbank  \& Richard}{Yuan
  et~al.}{2012}]{Yuan2012TheRedshift}
Yuan T.~T.,  Kewley L.~J.,  Swinbank A.~M.,   Richard J.,  2012, \mn@doi [The
  Astrophysical Journal] {10.1088/0004-637X/759/1/66}, 759, 66

\bibitem[\protect\citeauthoryear{{Zakamska} et~al.,}{{Zakamska}
  et~al.}{2016}]{zak16}
{Zakamska} N.~L.,  et~al., 2016, \mn@doi [Monthly Notices of the Royal
  Astronomical Society] {10.1093/mnras/stv2571}, \href
  {https://ui.adsabs.harvard.edu/abs/2016MNRAS.455.4191Z} {455, 4191}

\bibitem[\protect\citeauthoryear{Zheng et~al.,}{Zheng
  et~al.}{2012}]{Zheng2012ABang}
Zheng W.,  et~al., 2012, \mn@doi [Nature] {10.1038/nature11446}, 489, 406

\makeatother
\end{thebibliography}


\bsp	
\label{lastpage}
\end{document}